\documentclass[12pt,a4paper]{article}

\usepackage{amsfonts}
\usepackage{amsmath}
\usepackage{amsbsy}
\usepackage{theorem}
\usepackage{graphicx}

\begin{document}
\title{Point pattern analysis and classification on compact two--point homogeneous spaces evolving time}
\author{M. P. Fr\'{\i}as$^1$, A. Torres$^{2}$ and  M. D. Ruiz--Medina$^{3}$}
\date{}

 \maketitle

\bigskip

\begin{abstract} This paper introduces a new modeling framework for the statistical analysis of point patterns on a manifold  $\mathbb{M}_{d},$ defined by a connected and compact two--point homogeneous space, including   the special case of the sphere.
 The presented  approach is based on  temporal Cox processes driven by a $L^{2}(\mathbb{M}_{d})$--valued log--intensity. Different   aggregation schemes on the manifold of the spatiotemporal  point--referenced data are implemented in terms of the time--varying  discrete   Jacobi polynomial transform of the log--risk process.  The  $n$--dimensional microscale point pattern evolution in time  at different manifold spatial scales is then characterized from such a transform.  The simulation study undertaken illustrates the construction of spherical point process models  displaying aggregation at low Legendre  polynomial transform  frequencies (large  scale), while regularity is observed at   high frequencies (small scale). $K$--function analysis  supports these results   under temporal  short--, intermediate-- and long--range dependence  of the log--risk process.
\end{abstract}

\noindent \emph{Keywords:}
Connected and compact two--point homogeneous spaces;  Cox processes;  discrete Jacobi polynomial transform;  $K$--function; $\mathbb{M}_{d}$--supported   random fields; point pattern analysis; statistical distances.



\section{Introduction}\label{sec1} Several  statistical approaches arise for processing spatial  areally-\linebreak aggregated  or/and misalignment data in several environmental disciplines  requiring, for example,  the application of Geophysical, Ecological and  Epidemiological models.  The approach presented in this paper goes beyond the Euclidean setting, analyzing count models on a manifold defined by a connected and compact two--point homogeneous space.    Under spatial  isotropy we consider   weighted aggregation schemes adapted to the geometry of the manifold, in terms of  the elements of the Jacobi polynomial  basis (see Theorems 4 and 5 in  \cite{MaMalyarenko}, and
 \cite {Marinucci}   for the special case of the sphere). The application of harmonic analysis in this more general context leads to the characterization of the evolution of point patterns at different spatial scales in the manifold.

\medskip

Markov random ﬁeld (MRF) models, particularly,  Conditional Autoregressive (CAR) models have been widely applied  to represent the dynamics of the log--intensity process, interpreted as a log--risk process in the  context of double stochastic Poisson processes, also named  Cox processes (see \cite{Besag }).
In disease mapping, areal disease   counts have been usually analyzed under this Markovian log--risk process  framework (see, e.g., \cite{Ugarte2009}; \cite{Ugarte2010a};   \cite{Ugarte2012}). Particularly, different parametric, semiparametric and nonparametric statistical approaches  have been adopted in the estimation of deterministic and   random intensities (see \cite{Baddeleyeta06};  \cite{Diggleetal10b}; \cite{GB18}; \cite{Guan06},  and the references therein).   In point pattern analysis, special attention has been paid to functional summary statistics like the nearest
neighbour, empty space,   and  $K$ functions  (see, e.g.,   \cite{Diggleb};\cite{Illian08}). Recently, LASSO estimation based on spherical autoregressive processes has been proposed in  \cite{CaponeraDurastanti} beyond the Euclidean setting.

\medskip

Alternatively, in the functional data analysis (FDA) framework,
conditional autoregressive Hilbertian process (CARH process) models were considered by \cite{Cugliari2011},   \cite{Cugliari2013} and \cite{Guillas 2002},
developing projection estimation methods for prediction.
 In \cite{RuizMedina14},  an Autoregressive Hilbertian process  (ARH(1) process) framework was adopted to represent the dynamics of the spatiotemporal log--risk process.  This framework has  also been   adopted in \cite{Torres21b} for COVID--19 mortality prediction by applying   multivariate curve regression and machine learning.
 As an alternative, to analyze the spatial interaction between log--risk curves at different regions,  in \cite{FriasTorres22},   a Spatial Autoregressive Hilbertian process (SARH(1)  process) based modeling was applied. Recently,  wavelet--based projection methods are implemented in  \cite{Torres21a}   to developing an infinite--dimensional  spatial multiresolution point pattern analysis, based on spatiotemporal  Log-Gaussian Cox processes in the Euclidean setting.  The present paper goes beyond this Euclidean setting. At each spatial resolution level on the manifold, defined in terms of time--varying discrete Jacobi transform, temporal point pattern analysis is achieved from the latent random  intensity process in time, and its  higher order moments. In the particular case of the sphere,  suitable log--intensity models can be found in \cite{CaponeraMarinucci}, where   spherical functional autoregressive (SPHAR) processes are introduced, and their  asymptotically analysis is derived. Additionally,  spherical functional autoregressive--moving average (SPHARMA) processes are considered in \cite{Caponera21}, extending SPHAR  processes, for suitable approximation of  isotropic and stationary sphere--cross--time random fields. Here,  functional spectral analysis tools are applied, and Wold--like decomposition results are  derived.

\medskip

    A growing interest on spherical  point processes, and its functional summary statistics  is observed in recent contributions  (see, e.g.,  \cite{MollerRubak16}; \cite{Robeson14}). In this paper, our interest relies on    point patterns analysis in compact two--point homogeneous spaces evolving time. The framework of  temporal Cox  processes driven by
 log--intensities, evaluated in the space $L^{2}(\mathbb{M}_{d}, d\nu)$  of square integrable functions on a compact two--point homogeneous space  $\mathbb{M}_{d}$ is then considered.  Particularly, $\mathbb{M}_{d}$ is a manifold with $d$ denoting its topological dimension, and $d\nu $ denotes its measure,  induced by the probabilistic invariant measure on  the connected component of the group of isometries of $\mathbb{M}_{d}.$
 The associated infinite--dimensional $n$--order product density is  identified with the infinite product of  temporal  $n$--order product densities. A spatial multi--scale analysis of   the point process  evolution is achieved from these  temporal  $n$--order product densities, and the usual functional  summary statistics constructed from them.

 The interest of the extended  family of Cox  processes analyzed here relies on well--known examples of compact two--point homogeneous spaces   like the sphere  $\mathbb{S}_{d}\subset\mathbb{R}^{d+1},$ and the   projective spaces over different algebras (see Section 2 in \cite{MaMalyarenko}  for more details). Recent advances on modeling, analysis and  simulation of Gaussian spherical isotropic random fields, including random fields obeying a fractional stochastic partial differential equation on the sphere, can be exploited in our more general   $L^{2}(\mathbb{M}_{d}, d\nu)$--valued  Gaussian log--risk   process  framework (see \cite{Alegria20}; \cite{Anh2018}; \cite{Emery19}; \cite{Leonenko21}, among others).  Particularly, \cite{Anh2018}  and \cite{Leonenko21} focalize on   Cosmic Microwave Background   (CMB) evolution modeling and data analysis.  The approach presented here can contribute to this  modeling framework to approximate  the distribution of CMB hot and cold spots.

\medskip

  In point pattern analysis on  a $d$--dimensional manifold $\mathbb{M}_{d},$ embedded into $\mathbb{R}^{d+1},$ one can apply
the isometric identification of  $(\mathbb{S}_{d}, d_{\mathbb{S}_{d}})$  with  $(\mathbb{M}_{d}, d_{\mathbb{M}_{d}})$  via the identity $d_{\mathbb{S}_{d}}(\mathbf{x}_{1},\mathbf{x}_{2})=\arccos \left(\mathbf{x}_{1}^{T}\mathbf{x}_{2}\right),$ for  $\mathbf{x}_{1},\mathbf{x}_{2}\in
\mathbb{S}_{d}.$
  \noindent This geodesic distance  $d_{\mathbb{M}_{d}}$  is involved in the definition of functional summary statistics  characterizing the aggregation, regularity or  inhibition of the point pattern. In particular, point pattern classification is achieved in terms of this geodesic distance.
This paper presents a  new manifold spatial--scale--dependent point pattern classification analysis over time, via time--varying discrete Jacobi transform, achieved  in terms of  different statistical distances. $K$ function  analysis is also performed describing the cumulative counting properties of pair correlation function in time through different spatial scales.  In the simulation study undertaken on the sphere, temporal  short--, intermediate-- and long--range dependence  models are tested, the statistical distance based methods implemented reflect a  departure from complete randomness of the point pattern   at coarser (large) scales in the manifold  (low frequencies  of the  time--varying discrete  Legendre polynomial transform). While their small scale (high--frequency) behavior shows regularity. $K$--function based analysis supports  the same classification results, independently of the underlying dependence range of the log--intensity. At coarser spatial  scales, stronger departure from point pattern regularity is observed  when long--range dependence log--intensity models are tested. As mentioned above,  the approach presented in this paper then   provides a framework to detect  non--uniformity of the spherical distribution of CMB hot and cold spots, since these deviations from uniformity  are usually geometrically described in terms of  clustering, girdling or ring structures (see, e.g.,  \cite{Khan21}; \cite{Sadr21}).

\medskip

The outline of the paper is the following. Preliminaries on connected and compact two--point homogeneous spaces are given in Section \ref{sec2}. The new class of Cox processes analyzed in a metric space framework is introduced in Section \ref{sec3}. The proposed statistical distance based classification methodology through  spatial scales, involving  $n$--order product density,  is formulated  in Section \ref{sec4a}.   $K$ function is also explicitly computed from the time--varying discrete Jacobi transform of the second--order structure of the $L^{2}(\mathbb{M}_{d})$--valued temporal  log--intensity.  The results of the simulation study undertaken are displayed in Section \ref{sec4b}. Some final remarks and discussion can be found in Section \ref{sec5} to ending the paper.

   \section{Preliminaries}\label{sec2}
Let $\{ X_{t}(\cdot ),\ t\in \mathcal{T}\subseteq \mathbb{R}\}$ be an  infinite--dimensional random process such that, for each $t\in \mathcal{T}\subseteq \mathbb{R},$ almost surely $\log(X_{t})\in L^{2}(\mathbb{M}_{d}),$  and \linebreak  $E[\log(X_{t})]\underset{ L^{2}(\mathbb{M}_{d})}{=}0,$ with $\log(X_{t})$ having characteristic functional
   \begin{eqnarray}
   f_{\log(X_{t})}(h)&=&\int_{L^{2}(\mathbb{M}_{d})}\exp\left(i\left\langle h, \log(x_{t}) \right\rangle_{L^{2}(\mathbb{M}_{d})}\right)\mu_{\log(X_{t})}(d\log(x_{t}))\nonumber\\
   &=& \exp\left(-\frac{\left\langle \mathcal{R}_{0}(h), h\right\rangle_{L^{2}(\mathbb{M}_{d})}}{2}\right),\quad h\in L^{2}(\mathbb{M}_{d}),
   \label{fc}
   \end{eqnarray}
   \noindent where $\mathcal{R}_{0}=E\left[\log(X_{t})\otimes \log(X_{t})\right]\in \mathcal{L}^{1}(L^{2}(\mathbb{M}_{d}))$ denotes the covariance operator of $\log(X_{t}),$ and  $\mathcal{L}^{1}(L^{2}(\mathbb{M}_{d}))$ is the space of trace or nuclear operators on $ L^{2}(\mathbb{M}_{d}).$
   Here,  $\mu_{\log(X_{t})}$ is the induced Gaussian measure by $\log(X_{t})$ on $(L^{2}\left(\mathbb{M}_{d}), \mathcal{B}(L^{2}(\mathbb{M}_{d}))\right),$  with  $\mathcal{B}(L^{2}(\mathbb{M}_{d}))$ being the $\sigma$--algebra generated by all cylindrical subsets of $L^{2}(\mathbb{M}_{d}).$      In the subsequent development, we will also assume that, for any $t,s\in \mathcal{T},$  \begin{equation}E\left[\log(X_{t})(\mathbf{z}) \log(X_{s})(\mathbf{y})\right]= r_{t-s}\left(d_{\mathbb{M}_{d}}(\mathbf{z},\mathbf{y})\right)=    \widetilde{r}  \left(d_{\mathbb{M}_{d}}(\mathbf{z},\mathbf{y}), t-s\right),\ \mathbf{z}, \mathbf{y}\in\mathbb{M}_{d} ,\label{to}\end{equation}  \noindent  i.e.,   stationarity in time and isotropy over $\mathbb{M}_{d}$ in the weak sense are assumed. Note that the covariance operator $\mathcal{R}_{t-s}$ with kernel $r_{t-s}(\cdot ,\cdot )$ is  a nuclear operator, and its kernel $r_{t-s}\left(d_{\mathbb{M}_{d}}(\mathbf{z},\mathbf{y})\right)$ is assumed to be continuous.

 For the special case  $r_{t-s}(\cdot ,\cdot )=r_{s-t}(\cdot ,\cdot ),$  the following series expansion is obtained from  Theorems 4 and 5 in  \cite{MaMalyarenko}:
\begin{eqnarray}
&&\log(X_{t})(\mathbf{z})=\sum_{n=0}^{\infty}V_{n}(t)P_{n}^{(\alpha, \beta )}\left(\cos \left(d_{\mathbb{M}_{d}}(\mathbf{z},\mathbf{U}\right)\right),\ \mathbf{z}\in \mathbb{M}_{d},\ t\in \mathbb{R},
\label{klexp}
\end{eqnarray}
\noindent where $P_{n}^{(\alpha, \beta )}$ is a Jacobi polynomial of degree $n$ depending on parameter vector $(\alpha , \beta )$  (see, e.g., \cite{Andrews99}). Here, $\{V_{n}(t),\ n\in \mathbb{N}_{0}\}$ is a sequence of independent stationary random processes on $\mathcal{T}\subseteq \mathbb{R},$  satisfying
$E[V_{n}(t)]=0$ and  $E[V_{n}(t_{1})V_{n}(t_{2})]=a_{n}^{2}b_{n}(t_{1}-t_{2}),$ $n\in \mathbb{N}_{0}.$
The random variable $\mathbf{U}$ is uniformly distributed on $\mathbb{M}_{d},$ and is independent of $\{V_{n}(t),\ n\in \mathbb{N}_{0}\},$  and $\sum_{n=0}^{\infty}b_{n}(0)P_{n}^{(\alpha,\beta )}(1)$ converges. Also,  $$\mbox{cov}\left(V_{n}(t)P_{n}^{(\alpha, \beta )}\left(\cos \left(d_{\mathbb{M}_{d}}(\mathbf{z},\mathbf{U}\right)\right),V_{m}(t)P_{m}^{(\alpha, \beta )}\left(\cos \left(d_{\mathbb{M}_{d}}(\mathbf{z},\mathbf{U}\right)\right) \right)=0,$$ \noindent for $m\neq n,$   $\mathbf{z}\in \mathbb{M}_{d},$  and
$t\in \mathcal{T}.$

\section{Cox processes family}\label{sec3}
Let now consider the measure $d\nu(\mathbf{z})$ induced on the homogeneous space $\mathbb{M}_{d}=G/K,$ by the probabilistic invariant measure on $G,$ with $G$ being the connected component of the group of isometries of $\mathbb{M}_{d},$ and $K$ be the stationary subgroup of a fixed point $\mathbf{o}\in \mathbb{M}_{d}.$ As before, $H=L^{2}\left(\mathbb{M}_{d},d\nu(\mathbf{z})\right).$

Our spatiotemporal   count data model $\{N_{t}(\cdot),\ t\in \mathcal{T}\}$   characterizes the behavior of the temporal family   $\mathbf{Y}=\{\mathbf{Y}_{t},\ t\in \mathcal{T}\subseteq \mathbb{R}\}$   of  finite  point sets of $\mathbb{M}_{d},$  randomly arising at different  times in the interval family $\{[0,t],\ t\in \mathcal{T}\}.$    Specifically,
for every $t\in \mathcal{T},$ and any Borel set $A\subseteq \mathbb{M}_{d},$  $N_{t}(A)$
denotes the number of points  in the pattern $\mathbf{Y}_{t}$ falling in the  region $A\subseteq \mathbb{M}_{d},$  randomly arising in the interval $[0,t].$   Here, we consider the $\sigma$--algebra $\mathcal{F}$ generated by the events $\{ N_{t}(A)=n\}$ indicating that $n$ points in $\mathbf{Y}_{t}$ are falling in a region $A\subseteq \mathbb{M}_{d},$ at some specific times in $[0,t],$    for any Borel set $A\subseteq \mathbb{M}_{d},$  interval $[0,t],$ and integer $n\in \mathbb{N}.$

Assume that $\{N_{t}(\cdot), \ t\in \mathcal{T}\}$ defines a spatiotemporal Cox process with random log--intensity $\log(X_{t}),$  whose infinite--dimensional  marginals have characteristic functional (\ref{fc}). The $n$--dimensional  micro--scale  behavior of the random point pattern
is then characterizes by its   $n$--order product density $\rho^{(n)}_{t_{1},\dots,t_{n}}(\mathbf{z}_{1},\dots, \mathbf{z}_{n}),$ with  $$\rho_{t_{1},\dots, t_{n}}^{(n)}(\mathbf{z}_{1},\dots,\mathbf{z}_{n})d\nu^{(n)}(\mathbf{z}_{1},\dots,\mathbf{z}_{n})dt_{1},\dots, dt_{n}$$\noindent  indicating  the probability that $\mathbf{Y}_{t}$ has a point in each of $n$ infinitesimally small regions  on $\mathbb{M}_{d}$ around $\mathbf{z}_{1},\dots,\mathbf{z}_{n},$
of surface measure $d\nu(\mathbf{z}_{1})\cdots d\nu(\mathbf{z}_{n}),$
 over the infinitesimal time intervals around $t_{1},\dots, t_{n},$ of length $dt_{1},\dots, dt_{n}.$   Under the modeling framework introduced in Section \ref{sec2}, from equation (\ref{klexp}),
  for any  $t_{1},\dots, t_{n}\in \mathbb{R},$
  one can compute $\rho_{t_{1},\dots,t_{n}}^{(n)}$ as follows:
\begin{eqnarray}
&&\rho^{(n)}_{t_{1},\dots,t_{n}}(\mathbf{z}_{1},\dots, \mathbf{z}_{n})=E\left[\prod_{i=1}^{n}\exp\left(X_{t_{i}}(\mathbf{z}_{i} )\right)\right]=E\left[\exp\left(\sum_{i=1}^{n}X_{t_{i}}(\mathbf{z}_{i} )\right)\right]\nonumber\\
&&=[\rho]^{n}\exp\left(\frac{1}{2}\sum_{i=1}^{n}\sum_{j=1}^{n}
\sum_{q=0}^{\infty}b_{q}(t_{i}-t_{j})P_{q}^{(\alpha, \beta )}\left(\cos\left(d_{\mathbb{M}_{d}}(\mathbf{z}_{i},\mathbf{z}_{j})\right)\right)
\right),\nonumber\\
  \label{kernelsncm3}\end{eqnarray}
\noindent  for every  $z_{i}\in \mathbb{M}_{d},$  $ i=1,\dots,n.$
 In particular, for any $t\in \mathcal{T},$ and, for any $t_{1},t_{2}\in \mathcal{T},$
the intensity function  $\rho_{t} = \rho_{0} =\rho^{(1)}(t),$  and the pair correlation function   $g_{t_{1}-t_{2}}\left(\cos\left(d_{\mathbb{M}_{d}}(\mathbf{z}_{1},\mathbf{z}_{2}) \right)\right),$ $\mathbf{z}_{1},\mathbf{z}_{2}\in \mathbb{M}_{d},$
respectively admit the following expressions:
\begin{eqnarray}&&
\rho=\rho_{0}(\mathbf{z}) =\exp\left(\frac{1}{2}\sum_{q=0}^{\infty}b_{q}(0)P_{q}^{(\alpha, \beta )}\left(1\right)\right)=
\prod_{q=1}^{\infty}\rho_{q}
,\ \forall \mathbf{z}\in \mathbb{M}_{d},\label{if}\\& &
g_{t_{1}-t_{2}}\left(\cos\left(d_{\mathbb{M}_{d}}(\mathbf{z}_{1},\mathbf{z}_{2}) \right)\right)= \frac{\rho^{(2)}_{t_{1}-t_{2}}\left(\cos\left(d_{\mathbb{M}_{d}}(\mathbf{z}_{1},\mathbf{z}_{2})\right)\right)}{\rho^{2}}\nonumber\\
& & =
\exp\left(\sum_{n=0}^{\infty}b_{n}(t_{1}-t_{2})P_{n}^{(\alpha, \beta )}\left(\cos\left(d_{\mathbb{M}_{d}}(\mathbf{z}_{1},\mathbf{z}_{2})\right)\right)\right).\label{pcfvf}
\end{eqnarray}

In our subsequent spatial multi--scale  temporal point pattern analysis on connected and  compact two--point homogeneous spaces, we apply  the identification of the  $n$--order product density $\rho^{(n)}_{t_{1},\dots,t_{n}}(\mathbf{z}_{1},\dots, \mathbf{z}_{n})$ in equation
(\ref{kernelsncm3}) with  the infinite product of temporal $n$--order product densities at different spatial resolution scales, defined from the discrete  Jacobi transform, i.e.,
\begin{eqnarray}
&&\hspace*{-1cm}\rho^{(n)}_{t_{1},\dots,t_{n}}(\mathbf{z}_{1},\dots, \mathbf{z}_{n})=[\rho]^{n}\exp\left(\frac{1}{2}\sum_{i=1}^{n}\sum_{j=1}^{n}
\sum_{q=0}^{\infty}b_{q}(t_{i}-t_{j})
P_{q}^{(\alpha, \beta )}\left(\cos\left(d_{\mathbb{M}_{d}}(\mathbf{z}_{i},\mathbf{z}_{j})\right)\right)
\right)\nonumber\\
&&=\prod_{q=0}^{\infty}[\rho_{q}]^{n}\exp\left(\frac{1}{2}\sum_{i=1}^{n}\sum_{j=1}^{n}b_{q}(t_{i}-t_{j})P_{q}^{(\alpha, \beta )}\left(\cos\left(d_{\mathbb{M}_{d}}(\mathbf{z}_{i},\mathbf{z}_{j})\right)\right)\right).
  \label{kernelsncm3smc}\end{eqnarray}
Thus, for each  $q\geq 1,$   \begin{eqnarray}&&\rho^{(n)}_{q}(t_{1},\dots,t_{n}, z_{1},\dots, z_{n})=[\rho_{q}]^{n}
\nonumber\\
&&\hspace*{1.5cm}\times
\exp\left(\frac{1}{2}\sum_{i=1}^{n}\sum_{j=1}^{n}b_{q}(t_{i}-t_{j})P_{q}^{(\alpha, \beta )}\left(\cos\left(d_{\mathbb{M}_{d}}(\mathbf{z}_{i},\mathbf{z}_{j})\right)\right)\right),\label{ndensityq}\end{eqnarray}
\noindent where the  Fourier coefficients $\{b_{q}(t_{i}-t_{j}),\ i,j=1,\dots,n\}$  characterize the behavior of  $n$--order product density at each  spatial scale $q\geq 0$ (see, e.g., Theorem 1.2.1 in \cite{Da Prato} where  infinite--dimensional Gaussian measures are identified with the infinite  product of one--dimensional measures).





\section{Point patterns classification through spherical scales}\label{sec4a}
Point pattern classification is performed  in this section by considering different   statistical  distances between $n$--order product densities at different manifold spatial scales. $K$-- function is  computed in terms of the time--varying discrete Jacobi transform of the second--order structure of the log--intensity or log--risk process.

We first consider the following Ibragimov contrast function, also known as Shannon--entropy--based  statistical  distance,  to measure the departure from complete randomness, by comparing   $n$--order product density (\ref{ndensityq})  with the $n$--order product density of homogeneous Poisson process on $\mathbb{M}_{d}$ evolving time  (see Section \ref{sec4b}  for its implementation):

\begin{eqnarray}&& D_{q}^{S}(\rho^{(n)}_{q},\rho_{q}^{n})
 =\int_{\mathcal{T}^{n}\times \mathbb{M}_{d}^{n}}\rho^{(n)}_{q}(t_{1},\dots,t_{n}, z_{1},\dots, z_{n})
\nonumber\\
&&\hspace*{0.5cm}\times
\ln \left( \frac{\rho^{(n)}_{q}(t_{1},\dots,t_{n}, z_{1},\dots, z_{n})}{[\rho_{q}]^{n}}\right)dt_{1}\cdots dt_{n}d\nu(z_{1}),\cdots, d\nu (z_{n})
\nonumber\\
&&=\int_{\mathcal{T}^{n}\times \mathbb{M}_{d}^{n}}[\rho_{q}]^{n}\exp\left(\frac{1}{2}\sum_{i=1}^{n}\sum_{j=1}^{n}b_{q}(t_{i}-t_{j})P_{q}^{(\alpha, \beta )}\left(\cos\left(d_{\mathbb{M}_{d}}(\mathbf{z}_{i},\mathbf{z}_{j})\right)\right)\right)
\nonumber\\
&&\hspace*{1.5cm} \times \frac{1}{2}\sum_{i=1}^{n}\sum_{j=1}^{n}b_{q}(t_{i}-t_{j})P_{q}^{(\alpha, \beta )}\left(\cos\left(d_{\mathbb{M}_{d}}(\mathbf{z}_{i},\mathbf{z}_{j})\right)\right)\nonumber\\
&&\hspace*{5cm}\times dt_{1}\cdots dt_{n}d\nu(z_{1}),\cdots, d\nu (z_{n}).
\nonumber\\
\label{midist}
\end{eqnarray}

Note that negative values of  $D_{q}^{S}(\rho^{(n)}_{q},\rho^{n}_{q})$ mean repulsiveness or inhibition at scale $q,$ while positive values mean aggregation, and null values correspond to the regular (complete  randomness) case at such a scale $q,$  in the $n$--order moment sense.
Ibragimov contrast function corresponds to the limiting case of a more general family of functions related to
R\'enyi--entropy based statistical   distances.  Specifically, one can consider for each $q\geq 0,$
\begin{eqnarray}&& D^{R}_{q,h}(\rho^{(n)}_{q},\rho_{q}^{n})
 =\frac{1}{h-1}\ln \left(\int_{\mathcal{T}^{n}\times \mathbb{M}_{d}^{n}}\rho^{(n)}_{q}(t_{1},\dots,t_{n}, z_{1},\dots, z_{n})\right.
\nonumber\\
&&\hspace*{0.5cm}\left.\times
\left[ \frac{\rho^{(n)}_{q}(t_{1},\dots,t_{n}, z_{1},\dots, z_{n})}{[\rho_{q}]^{n}}\right]^{h-1}dt_{1}\cdots dt_{n}d\nu(z_{1}),\cdots, d\nu (z_{n})\right)
\nonumber\\
&&\hspace*{-0.6cm}=\frac{1}{h-1}\ln \left(\int_{\mathcal{T}^{n}\times \mathbb{M}_{d}^{n}}[\rho_{q}]^{n}\exp\left(\frac{1}{2}\sum_{i=1}^{n}\sum_{j=1}^{n}b_{q}(t_{i}-t_{j})P_{q}^{(\alpha, \beta )}\left(\cos\left(d_{\mathbb{M}_{d}}(\mathbf{z}_{i},\mathbf{z}_{j})\right)\right)\right)\right.
\nonumber\\
&&\hspace*{1.5cm} \left.\times \exp\left(\frac{h-1}{2}\sum_{i=1}^{n}\sum_{j=1}^{n}b_{q}(t_{i}-t_{j})P_{q}^{(\alpha, \beta )}\left(\cos\left(d_{\mathbb{M}_{d}}(\mathbf{z}_{i},\mathbf{z}_{j})\right)\right)\right)\right.\nonumber\\
&&\hspace*{5cm}\left.\times dt_{1}\cdots dt_{n}d\nu(z_{1}),\cdots, d\nu (z_{n})\right),
\nonumber\\
\label{midistb}
\end{eqnarray}
\noindent where the  continuous positive shape parameter $h$ characterizes  the \linebreak $L^{h-1}(\mathcal{T}^{n}\times \mathbb{M}_{d}^{n},\rho^{(n)}_{q}(t_{1},\dots,t_{n}, z_{1},\dots, z_{n}),dt_{1}\cdots dt_{n}d\nu(z_{1}),\cdots, d\nu (z_{n}))$ space, whose norm is involved in measuring the aggregation or inhibition level of the point pattern at the logarithmic scale.
\subsection{Functional Summary  statistics}
\label{fss} In the Log--Gaussian Cox process framework,  the most  interesting case in
equation (\ref{midist}) corresponds to  $n=2,$ where one can alternatively compute the cumulative distribution function associated with the two--order product density, in terms of the pair correlation function  (\ref{pcfvf}), under stationarity in time and isotropy in space.  That is, we consider the following functional summary statistics  $K_{t}(\theta )$  under the assumption that $\mathbf{Y}$ is fully observed:
\begin{eqnarray}&&K_{t}(\theta )=\frac{1}{\rho^{2}|\mathcal{T}|\nu(\mathbb{M}_{d})}E\left[\sum_{(s,\mathbf{y})\in \mathbf{Y}}\sum_{(u,\mathbf{z})\in \mathbf{Y}\backslash \{(s,\mathbf{y})\}}1_{\{d_{\mathbb{M}_{d}}(\mathbf{z},\mathbf{y})\leq \theta \}}\otimes 1_{\{|s-u|\leq t \}}\right]\nonumber\\
&&=\frac{1}{\rho^{2}|\mathcal{T}|\nu(\mathbb{M}_{d})}\int_{\mathcal{T}^{2}\times \mathbb{M}_{d}^{2}}1_{\{d_{\mathbb{M}_{d}}(\mathbf{z},\mathbf{y})\leq \theta \}}(\mathbf{y},\mathbf{z})1_{\{|s-u|\leq t \}}(s,u)\rho_{s, u}^{(2)}(\mathbf{y},\mathbf{z})\nonumber\\
&&\hspace*{4cm}\times d\nu(\mathbf{y})d\nu(\mathbf{z})dsdu\nonumber\\
&&=\frac{1}{|\mathcal{T}|\nu(\mathbb{M}_{d})}\int_{\mathcal{T}^{2}\times \mathbb{M}_{d}^{2}}1_{\{d_{\mathbb{M}_{d}}(\mathbf{z},\mathbf{y})\leq \theta \}}(\mathbf{y},\mathbf{z})1_{\{|s-u|\leq t \}}(s,u)\nonumber\\
&&\hspace*{3.5cm}\times g_{s-u}(\cos \left(d_{\mathbb{M}_{d}}(\mathbf{y},\mathbf{z})\right))d\nu(\mathbf{y})d\nu(\mathbf{z})dsdu\nonumber\\
&&=\frac{1}{|\mathcal{T}|\nu(\mathbb{M}_{d})}\int_{\mathcal{T}^{2}\times \mathbb{M}_{d}^{2}}1_{\{d_{\mathbb{M}_{d}}(\mathbf{z},\mathbf{y})\leq \theta \}}(\mathbf{y},\mathbf{z})1_{\{|s-u|\leq t \}}(s, u)\nonumber\\
&&\hspace*{3.5cm}\times \exp\left(\sum_{q=0}^{\infty}b_{q}(s-u)P_{q}^{(\alpha, \beta )}\left(\cos\left(d_{\mathbb{M}_{d}}(\mathbf{y},\mathbf{z})\right)\right)
\right)\nonumber\\
&&\hspace*{4cm}\times
d\nu(\mathbf{y})d\nu(\mathbf{z})dsdu.
\label{kfunction}
\end{eqnarray}

Specifically, $K_{t}(\theta )$ function provides  the mean number of further points within geodesic distance $\theta$ occurring in a temporal interval of length less or equal than $t.$  For each spatial resolution $q\geq 1,$ we consider, for $0\leq \theta \leq \pi,$ and $t>0,$  the $K_{q}(t,\theta )$ function given by
\begin{eqnarray}
K_{q}(t,\theta ) &=&\frac{1}{|\mathcal{T}|\nu(\mathbb{M}_{d})}\int_{\mathcal{T}^{2}\times \mathbb{M}_{d}^{2}}1_{\{d_{\mathbb{M}_{d}}(\mathbf{z},\mathbf{y})\leq \theta \}}(\mathbf{y},\mathbf{z})1_{\{|s-u|\leq t \}}(s,u)\nonumber\\
&&\hspace*{0.5cm}\times \exp\left( b_{q}(s-u)P_{q}^{(\alpha, \beta )}\left(\cos\left(d_{\mathbb{M}_{d}}(\mathbf{y},\mathbf{z})\right)\right)
\right)\nonumber\\
&&\hspace*{3cm}\times
d\nu(\mathbf{y})d\nu(\mathbf{z})dsdu.
\label{kqfunctsrq}
\end{eqnarray}

\noindent At different spatial resolution levels $q,$  point pattern classification is achieved by comparing
 function $K_{q}(t,\theta )$ with   $K_{\mbox{Pois}}(t,\theta )=2t\pi (1-\cos(\theta )).$  The last one   corresponds  to complete randomness.
 Hence, one can respectively  interpret aggregation  and  inhibition at  spatial scale $q,$ when  $K_{q}(t,\theta )-K_{\mbox{Pois}}(t,\theta )>0,$   and  $K_{q}(t,\theta )-K_{\mbox{Pois}}(t,\theta )<0$ almost surely in $t$ and $\theta.$  The pointwise null values of this difference function  $K_{q}(t,\theta )-K_{\mbox{Pois}}(t,\theta )$ correspond to complete randomness. Specifically, one can compare $K_{q}(t,\theta )$  and $K_{\mbox{Pois}}(t,\theta )$ functions in terms of the $L^{p}$ norm of the quotient $K_{q}(t,\theta )/K_{\mbox{Pois}}(t,\theta )$ at logarithmic scale. On the other hand, pointwise information of the difference $K_{q}(t,\theta )-K_{\mbox{Pois}}(t,\theta ),$ for small and large temporal $t$ and angular $\theta $ distance arguments, respectively reflects  the small--scale and large--scale behavior of $K$--function. These behaviors are affected by the dependence range of the log--intensity process at coarser Jacobi spatial scales.  While they are almost invariant at higher resolution levels of the time--varying discrete Jacobi transform,  as given in Section  \ref{sec4b}  (see Figures \ref{fig3b}--\ref{fig3c}).

 For each $t>0,$ and $\theta \in [0,\pi],$  the nearest neighbour function $G_{t}(\theta )$ indicates    the mean number of points at a specific temporal $t,$ and angular  $\theta $ distances to the pattern.  The computation of this function requires the consideration of the  intensity function $\rho$ identified with the infinite product of uniform intensity functions $\rho_{q}$ at different spatial resolution scales $q$ in (\ref{if}),  which are constants under isotropy in space and stationarity in time, i.e.,
\begin{eqnarray}&&G_{t}(\theta )=\frac{1}{\rho |\mathcal{T}|\nu(\mathbb{M}_{d})}E\left[\sum_{(s,\mathbf{y})\in \mathbf{Y}}1_{\{\inf_{(\cdot,\mathbf{z})\in \mathbf{Y}\backslash \{(s,\mathbf{y})\}}d_{\mathbb{M}_{d}}(\mathbf{y}, \mathbf{z})\leq \theta \}}\right.\nonumber\\
&&\hspace*{5.5cm}\left.\otimes 1_{\{\inf_{(u,\cdot)\in \mathbf{Y}\backslash \{(s,\mathbf{y})\}} |s-u|\leq t \}}\right]= \frac{1}{|\mathcal{T}|\nu(\mathbb{M}_{d})}\nonumber\\
&&\hspace*{0.5cm}\times \int_{\mathcal{T}\times \mathbb{M}_{d}}1_{\{\inf_{(\cdot,\mathbf{z})\in \mathbf{Y}\backslash \{(s,\mathbf{y})\}}d_{\mathbb{M}_{d}}(\mathbf{y}, \mathbf{z})\leq \theta \}}(\mathbf{y}) 1_{\{\inf_{(u,\cdot)\in \mathbf{Y}\backslash \{(s,\mathbf{y})\}} |s-u|\leq t \}}(s)d\nu(\mathbf{y})ds.\nonumber\\
\label{nnf}
\end{eqnarray}
\noindent Its empirical counterpart is given, for $t\in \mathcal{T},$ and $\theta \in [0,\pi],$  by
\begin{eqnarray}\widehat{G}_{t}(\theta )&=&\frac{1}{N(\mathcal{T}\times \mathbb{M}_{d})}\sum_{(s,\mathbf{y})\in \mathbf{Y}}1_{\{\inf_{(\cdot,\mathbf{z})\in \mathbf{Y}\backslash \{(s,\mathbf{y})\}}d_{\mathbb{M}_{d}}(\mathbf{y}, \mathbf{z})\leq \theta \}}\nonumber\\ &&\hspace*{3.5cm}\otimes 1_{\{\inf_{(u,\cdot)\in \mathbf{Y}\backslash \{(s,\mathbf{y})\}} |s-u|\leq t \}},\nonumber\\\label{nnfev}\end{eqnarray}
\noindent provided that $N(\mathcal{T}\times \mathbb{M}_{d})=N_{\mathcal{T}}(\mathbb{M}_{d})>0.$
Given the stationarity and isotropy of the model considered,  the null values of  $D_{q}^{S}(\rho^{(n)}_{q},\rho_{q}^{n})$ for  $n=1,$  at every  scale $q\geq 0,$ in  equation (\ref{midist}),  excludes this  functional    summary statistics, $G_{t},$  for
   classification purposes.   The simulation study undertaken in the next section  illustrates  the global characterization of the      point pattern through the two--order product densities at different spatial scales, in terms of  statistical distances  $D_{q}^{S}(\rho^{(2)}_{q},\rho_{q}^{2}),$  $D_{q,h}^{R}(\rho^{(2)}_{q},\rho_{q}^{2}),$ and  $K$--function analysis from equations  in (\ref{midist}), (\ref{midistb})  and (\ref{kfunction}), respectively. This assertion is validated by computing  $D_{q}^{S}(\rho^{(3)}_{q},\rho_{q}^{3}),$ $q\geq 0,$ involving  third--order product densities.

\section{Simulation}\label{sec4b} 
\begin{figure}[!htb]
\begin{center}
\centerline{\includegraphics[width=10cm]{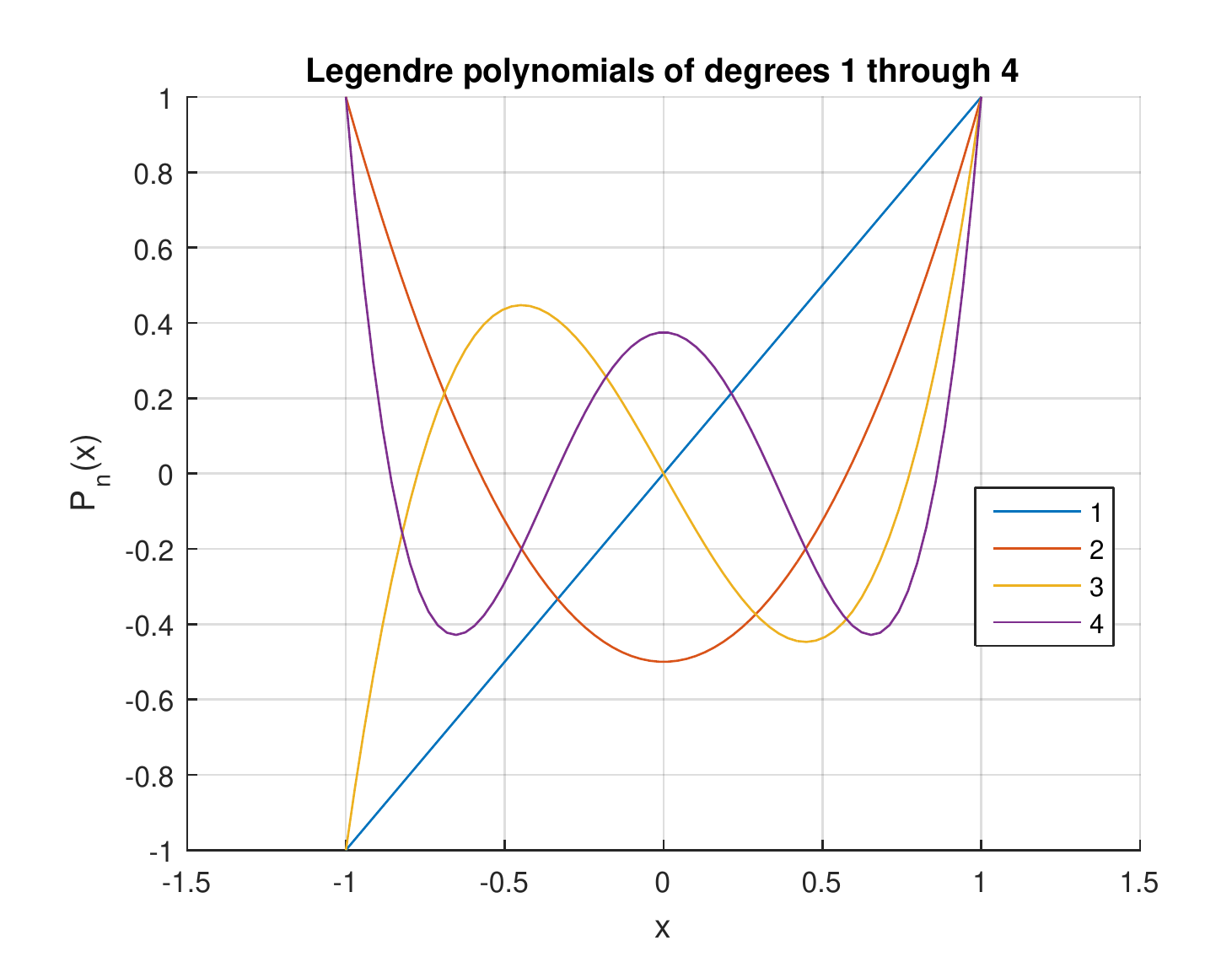}}
\caption{Legendre polynomial are plotted for orders $1,2,3,4.$ }\label{fig1a}
\end{center}
\end{figure}

In this simulation study, we restrict our attention to the case  of a Log--Gaussian Cox process on  $\mathbb{S}_{2}$ over the temporal interval $[0,10].$  For this special case, we work with the time--varying discrete Legendre transform, providing spherical large and small scale information about the log--intensity and its second--order structure by projection into the Legendre polynomials  $\{P_{l}\}$ (see Figure \ref{fig1a}, for $1,2,3,4$).

The  following parametric model is considered for the temporal covariance function of the Fourier random coefficients $\{V_{l}\}$ of the log--intensity $\log(X_{t})$ in equation (\ref{klexp}), with respect to the Legendre polynomial basis (see, e.g., \cite{CaponeraMarinucci};\cite{MarinucciRV}):
\begin{eqnarray}
   B_{l}(t,s)&=&E[V_{l}(t)V_{l}(s)]=(1/2)\frac{(l+1)^{-2-|t-s|}}{(1+(t-s)^{2})^{\theta\beta(l)}}  \nonumber\\
 \beta(l)&=&((8/10)(l+1))/((l+1)^{2}+1)^{1/2},\ l\geq 0,\ t,s \in \mathcal{T}.
        \label{temcovcoef}
        \end{eqnarray}

\begin{figure}[!htb]
\begin{center}
\centerline{\includegraphics[width=14cm]{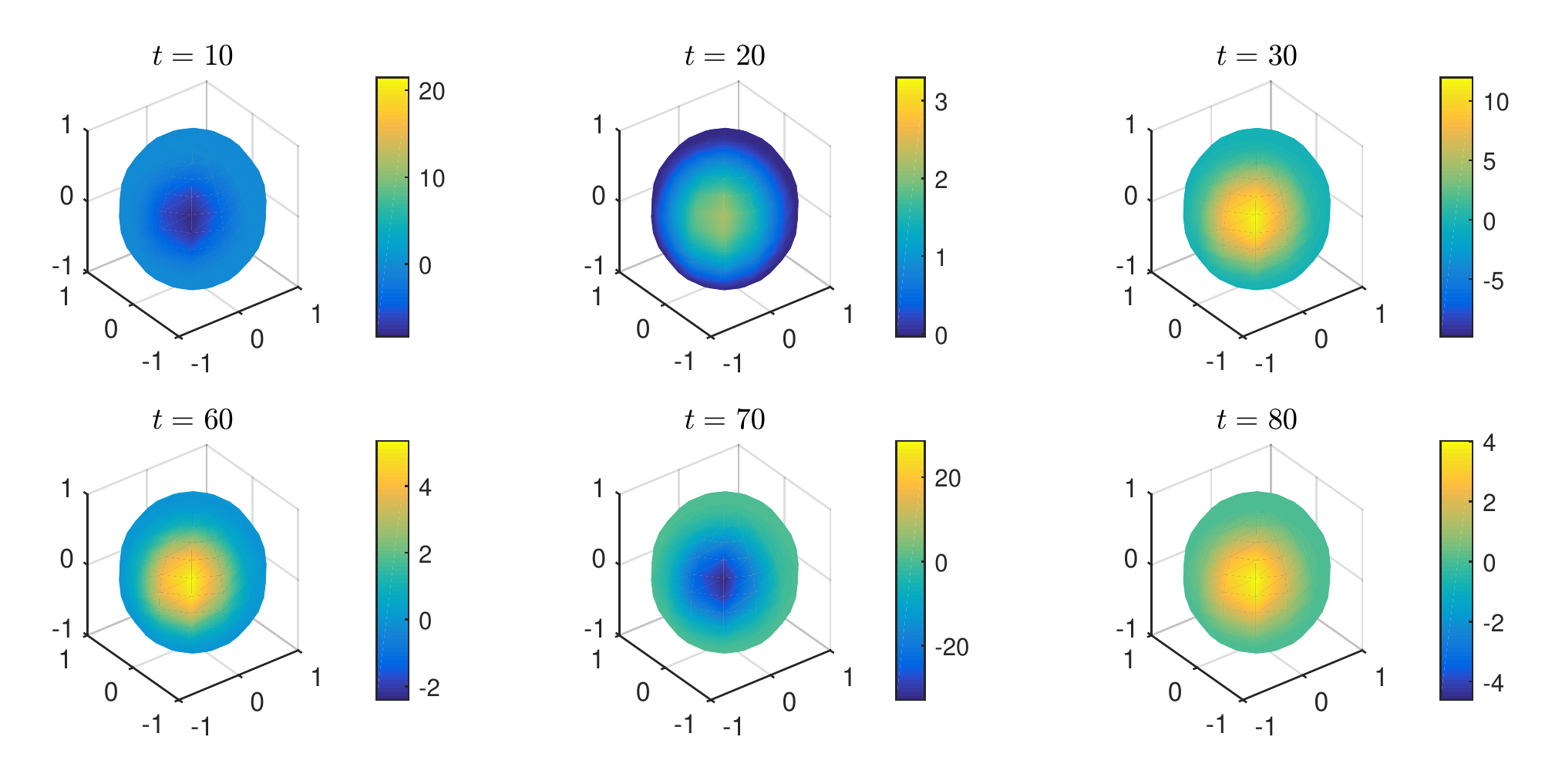}}
\caption{ \scriptsize{Log-intensity values on sphere for times $t=10,20,30,60,70,80.$
}}\label{fig1}
\end{center}
\end{figure}

        Thus, as given in Theorems 4  in  \cite{MaMalyarenko}, from (\ref{temcovcoef}), the kernel family \linebreak  $\{ r_{t-s}(\cdot,\cdot),\ t,s\in \mathcal{T}\}$ associated with the cross-covariance  operator family $\{ R_{t-s}=E[\log(X_{t})\otimes \log(X_{s})],\ t,s\in \mathcal{T}\}$   of the $L^{2}(\mathbb{S}_{2})$--valued log--intensity is given by:
        \begin{equation}r_{t-s}(\langle \mathbf{x}, \mathbf{y} \rangle) = \sum_{l=0}^{\infty} B_{l}(t-s)\frac{2 l + 1}{4\pi}P_{l}(\langle \mathbf{x}, \mathbf{y}\rangle),\ t,s\in \mathcal{T},\ \mathbf{x}, \mathbf{y}\in \mathbb{S}_{2}.\label{tcov}\end{equation}\noindent   Figure \ref{fig1} displays the values
 of the  Log--Gaussian intensity $\log (X_{t})$ in equation (\ref{klexp}), having covariance kernel (\ref{tcov}) for
 $\theta =1$ in (\ref{temcovcoef}), after truncating series expansion (\ref{klexp}) at  $M=5.$ In practice,
      model (\ref{temcovcoef}) is parametrically fitted by least--squares from a temporal $100\times 100$ regular grid,  from the projection into the Legendre basis of the empirical cross--covariance operators of the  data, whose functional values are approximated over a spherical regular grid of $225\times 225$ nodes.

Shannon--entropy  based distance  $D^{S}_{q}$ in  (\ref{midist}) is approximated by $\widehat{D}^{S}_{q}$ at  Legendre  scales $q=0,\dots, 30,$ to measure the statistical distance between the two--order product densities of the generated spherical Log--Gaussian Cox process, and the spherical homogeneous Poisson process over the interval $[0,10].$ The estimate $\widehat{D}^{S}_{q}$ is computed by applying Monte Carlo numerical integration, based on a sample  of size $1000,$ and least--squares parametric  five--degree polynomial  fitting for interpolation and smoothing.
Figure \ref{fig2} below displays three plots representing  the  values of $\widehat{D}^{S}_{q},$ for three embedded spatial  scale sets, i.e., for  $q$--values: $q=0,1,2,3,4,5$  (left--hand side),  $q=0,\dots,20$ (center) and $q=0, \dots, 30$ (right--hand side).
One can observe the positive values of the computed  statistical distances at Legendre scales zero to four indicating clustering, while null values are displayed from scales five to thirty. Maximum distance or aggregation level is attained at Legendre scales zero to one, decreasing to  zero distance through  scales two to four, leading to a regular behavior at Legendre  high frequencies ($q\in \{5, \dots, 30\}$), i.e.,  complete randomness at small scale.
\begin{figure}[!htb]
\begin{center}
\includegraphics[height=4cm, width=4cm]{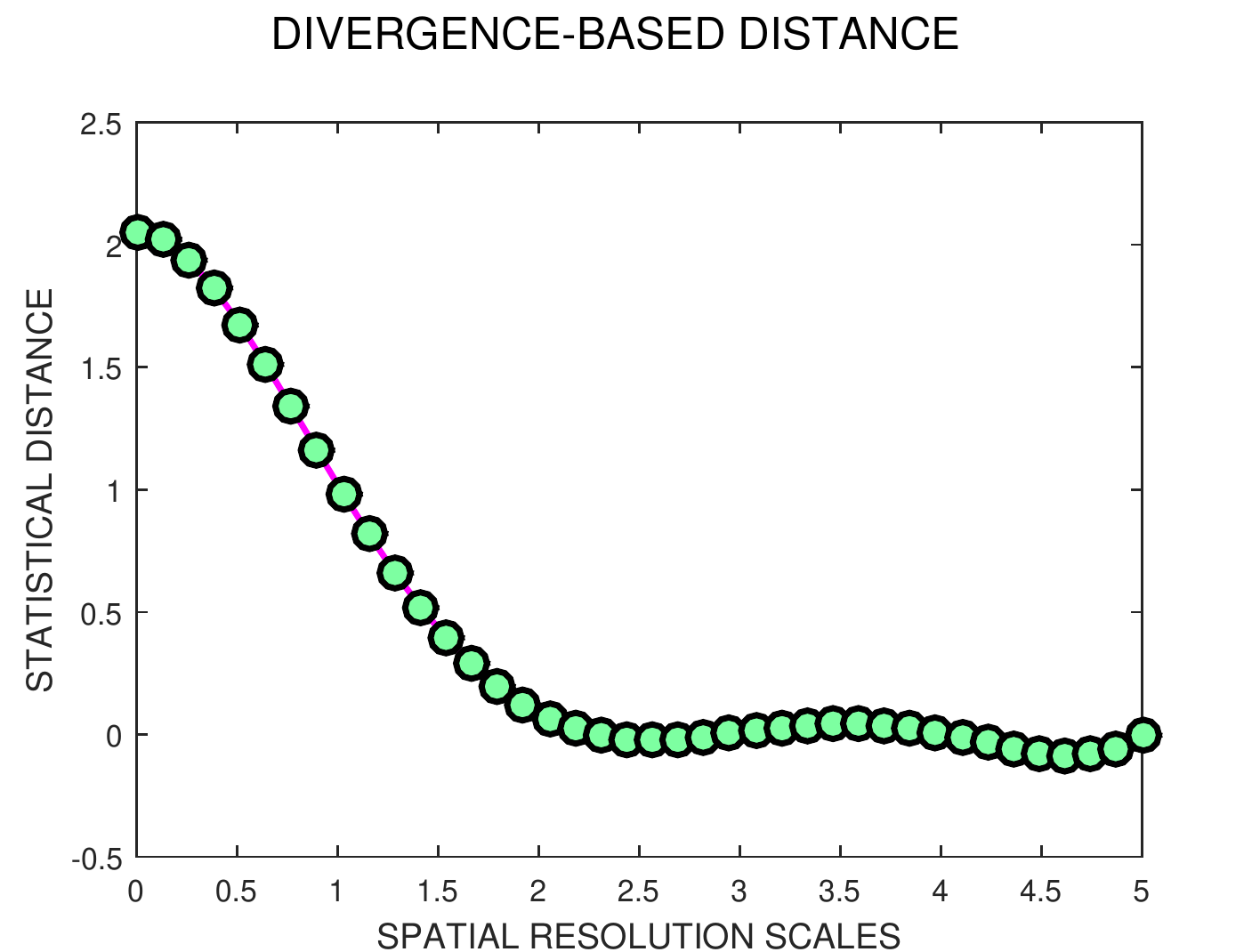}
\includegraphics[height=4cm, width=4cm]{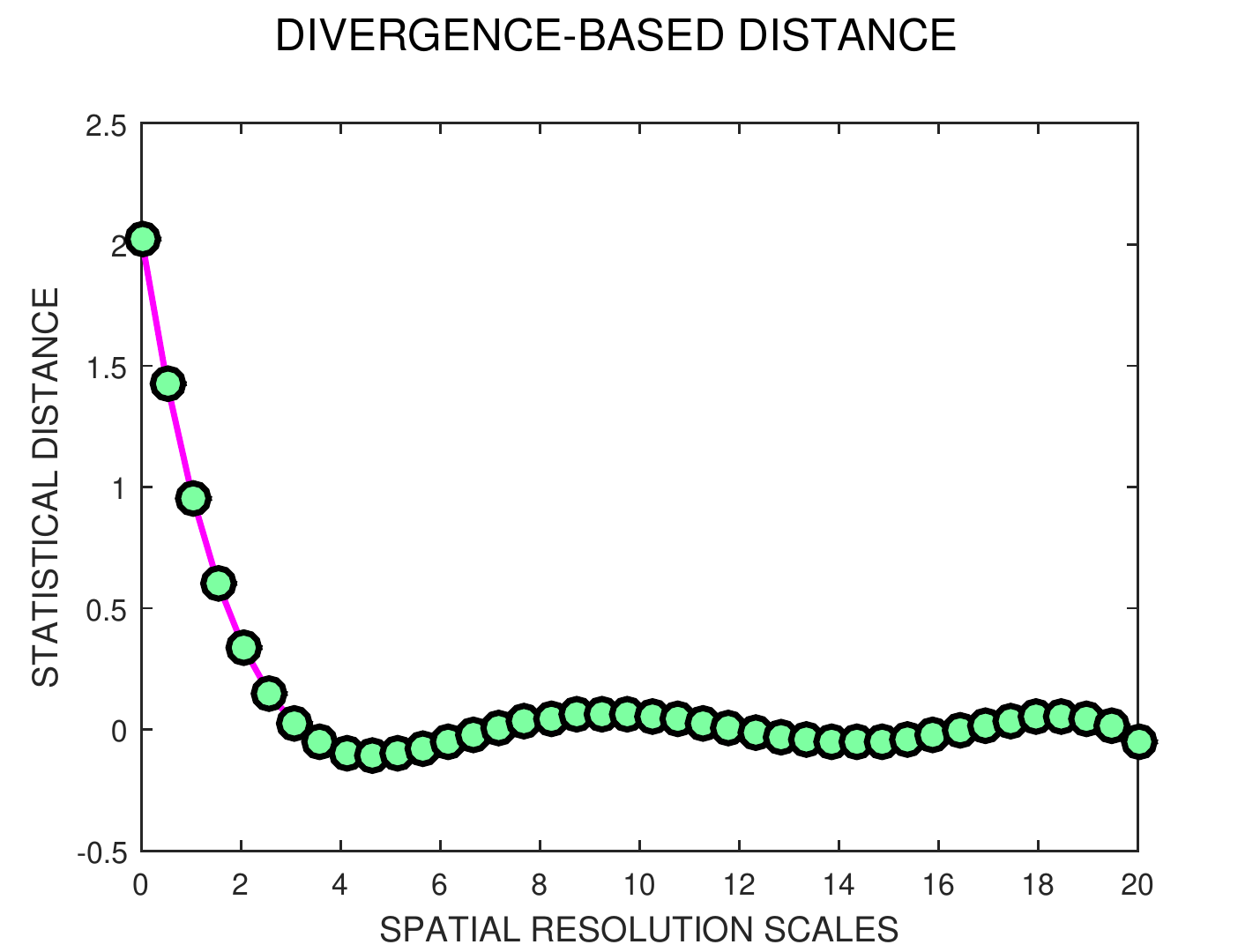}
\includegraphics[height=4cm, width=4cm]{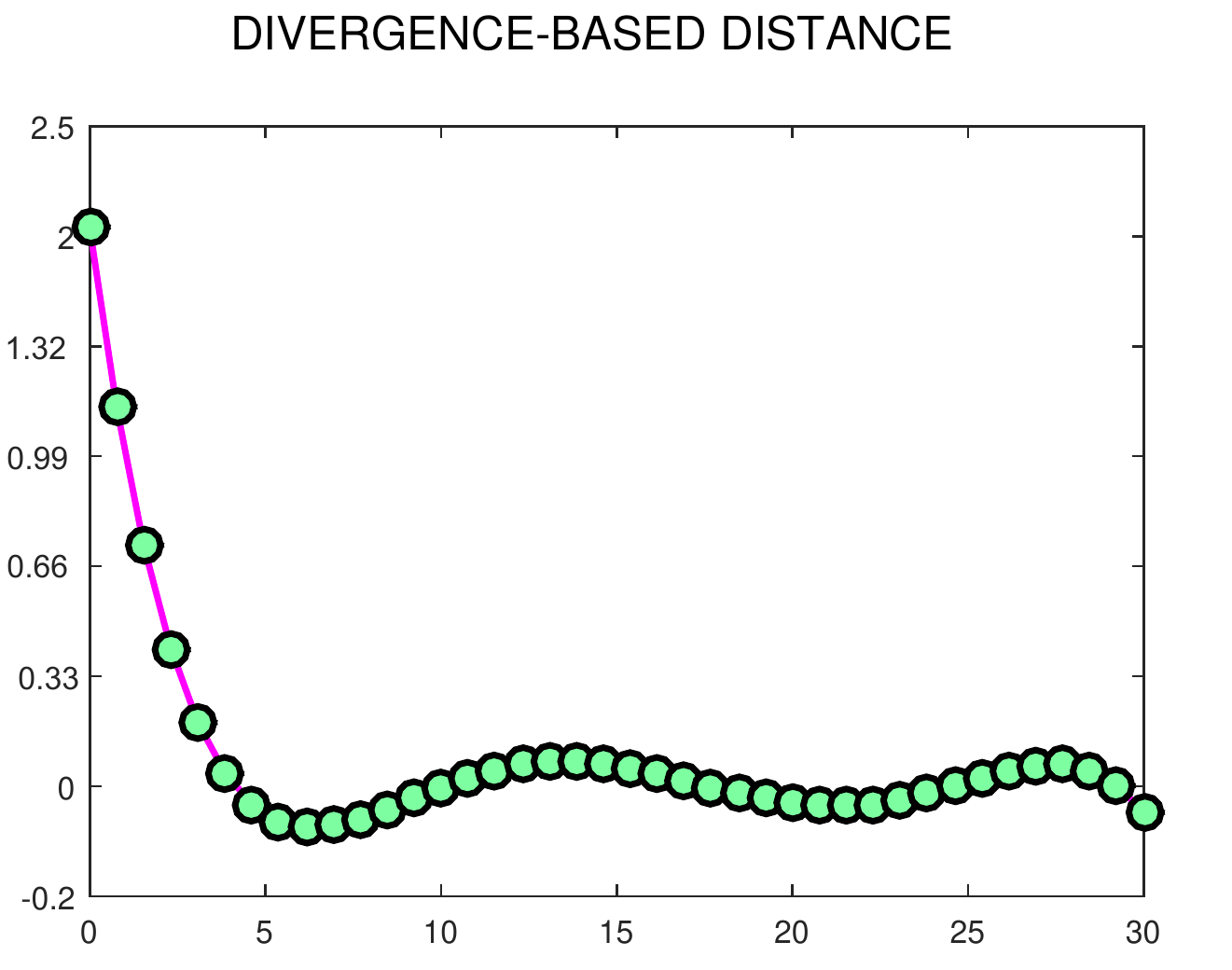}
\caption{ \scriptsize{Statistical distance based on Ibragimov contrast function  (\ref{midist}) between the two--order product  densities of the generated  spherical Log--Gaussian Cox process ($\theta =1$), and of spherical  homogeneous Poisson process  over the interval $[0,10],$ considering  Legendre  scales $q=0,1,2,3,4,5$ (left), $q=0,1,2,3,\dots, 20$ (center), and $q=0,1,2,3,\dots,30$ (right), reflected at the horizontal axis. }}\label{fig2}
\end{center}
\end{figure}

\begin{figure}[!htb]
\begin{center}
\includegraphics[height=8cm, width=8cm]{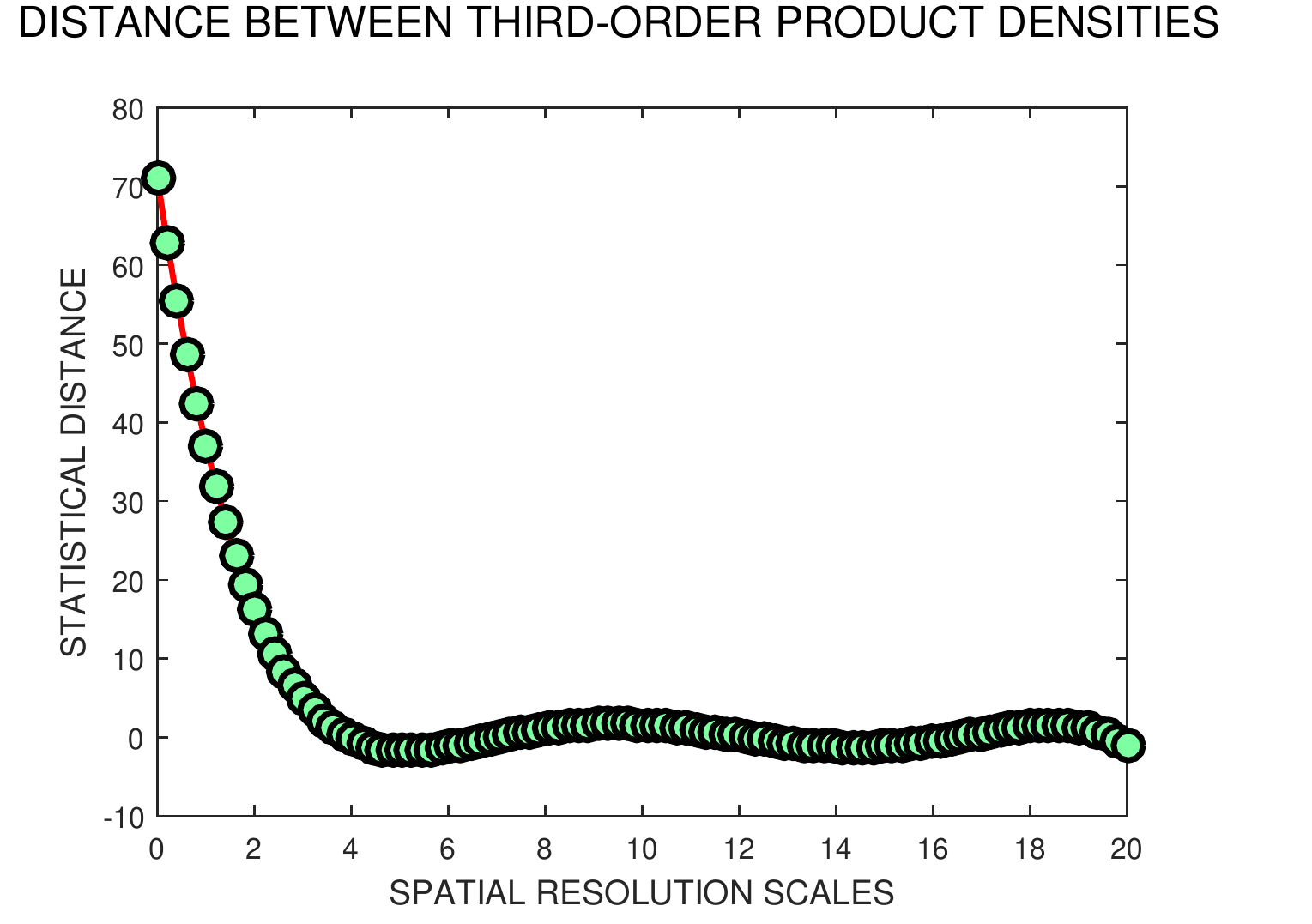}
\caption{ \scriptsize{Shannon--based  distance (\ref{midist}) for comparing third--order product densities of spherical  homogeneous Poisson process, and  spherical Log-Gaussian Cox process over the interval $[0,10].$ Here, the random coefficients $V_{n}$ in series expansion (\ref{klexp}) have covariance function (\ref{temcovcoef}) with $\theta =1.$}}\label{figTOD}
\end{center}
\end{figure}

Integral (\ref{midist}) defining $D_{q}^{S}(\rho^{(n)}_{q},\rho_{q}^{n})$ is computed for   $n=3$   by applying trapezoidal rule. As expected,
the classification results displayed in Figure \ref{fig2}  for the case of $n=2$ are supported
in the Log--Gaussian case for $n=3,$ over all Legendre scales  tested (see Figure \ref{figTOD}).

\begin{figure}[!htb]
\begin{center}
\includegraphics[height=5.5cm, width=12cm]{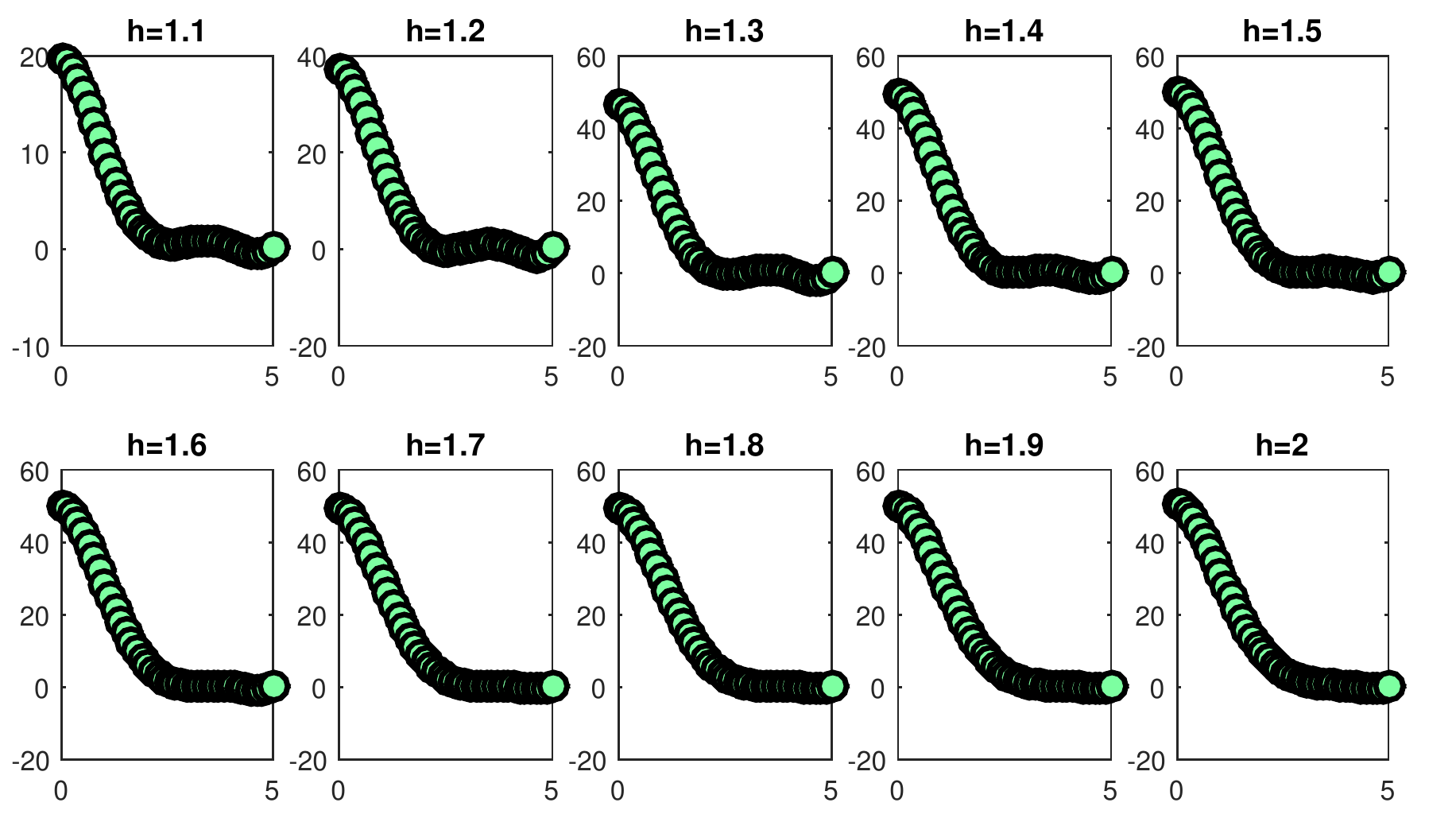}

\includegraphics[height=5.5cm, width=12cm]{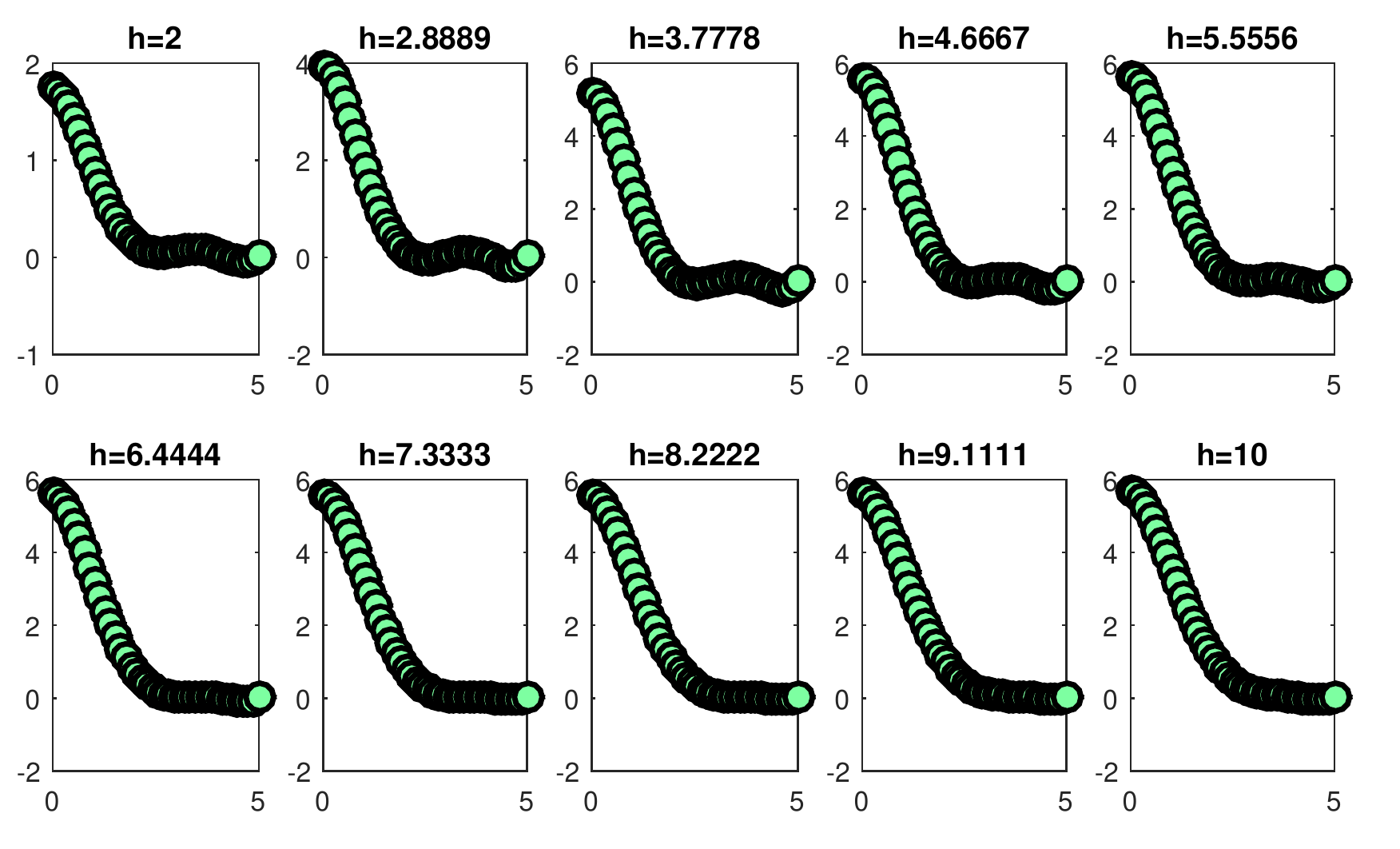}

\caption{ \scriptsize{Weak--dependent case ($\theta =100$). R\'enyi distances  $D_{q,h}^{R},$ $q=1,2,3,4,5$ (horizontal axis), and  $h\in (1,10).$ }}\label{fig2rd}
\end{center}
\end{figure}

\begin{figure}[!htb]
\begin{center}
\includegraphics[height=5.5cm, width=12cm]{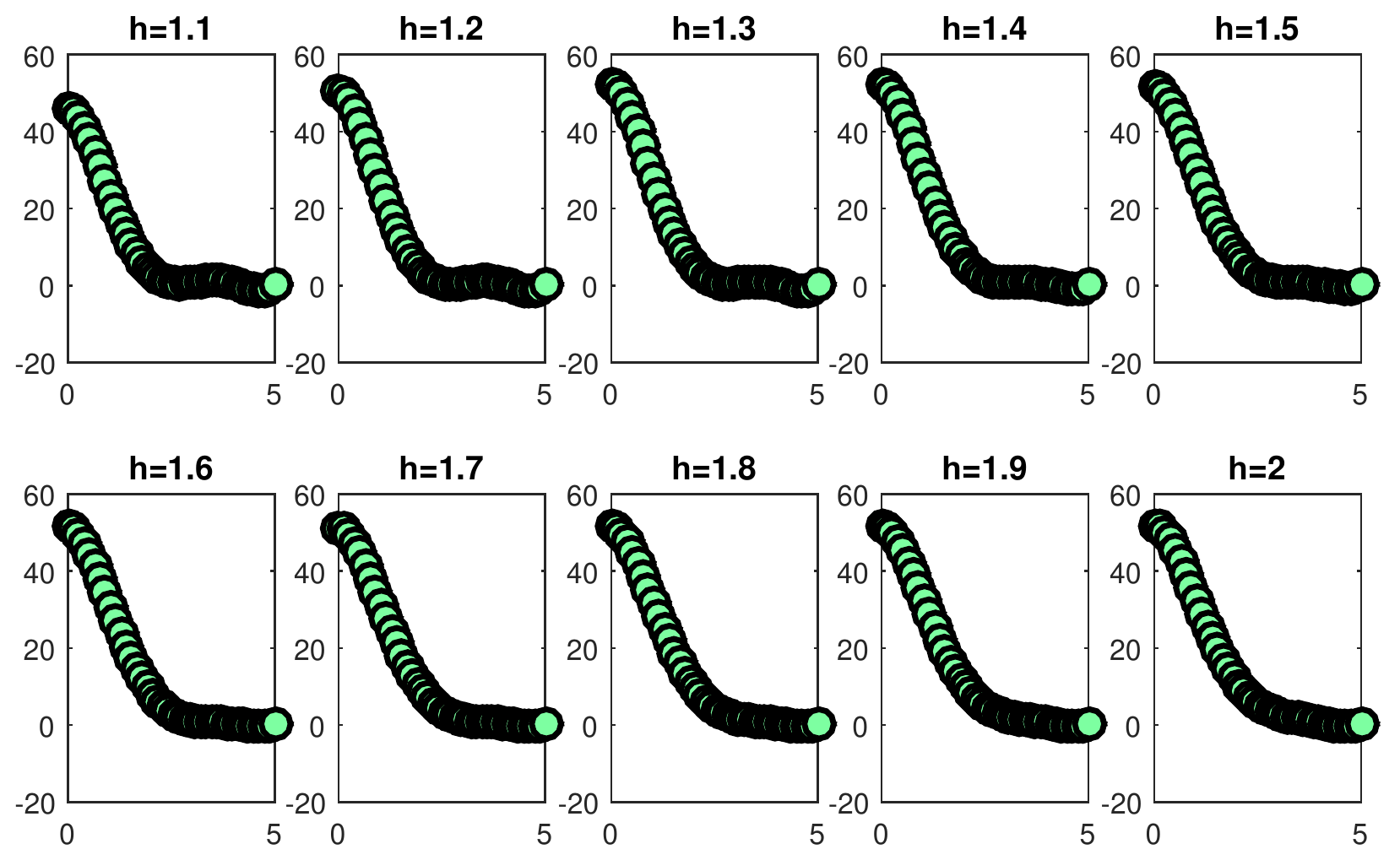}

\includegraphics[height=5.5cm, width=12cm]{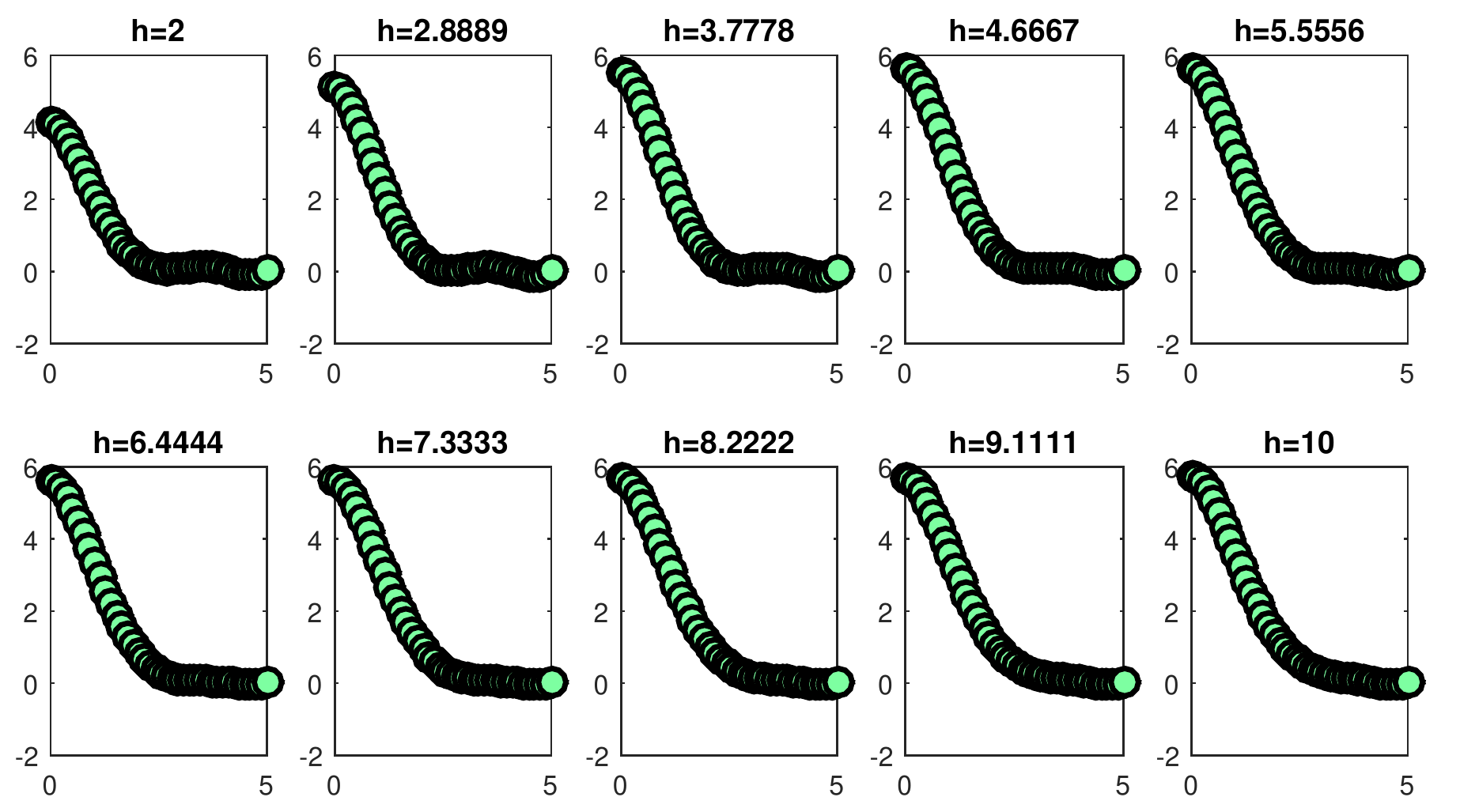}
\caption{ \scriptsize{Strong--dependent case ($\theta =1/100$). R\'enyi distances $D_{q,h}^{R},$  $q=1,2,3,4,5$  (horizontal axis), and     $h\in (1,10).$  }}\label{fig2rde}
\end{center}
\end{figure}

 Distance $D_{q,h}^{R}$ in (\ref{midistb}) is  now  approximated by $\widehat{D}^{R}_{q,h},$ computed by applying Monte Carlo numerical integration and five degree polynomial least--squares  smoothing.
A similar pattern to the one displayed at the left--hand--side plot in Figure \ref{fig2} is observed for the computed estimates of R\'enyi--entropy based distances $\widehat{D}_{q,h}^{R}$ of different integer and fractional orders $h,$   considering   Legendre scales $q=1,2,3,4,5.$  Such empirical distances provide additional information about  the  clustering index in the spatiotemporal point pattern. Figures \ref{fig2rd} and \ref{fig2rde} show such distances in the respective cases of short-- and long-- range dependence in time  of the log--intensity, corresponding to the values $\theta =100$ and  $\theta =1/100$ in equation (\ref{temcovcoef}).

\clearpage

All computed statistical distances reflect the same pattern with respect to Legendre scales   (horizontal axis), indicating regularity at Legendre scales larger or equal than five ($q\geq 5$), and clustering at Legendre scales  zero  to four ($q=0,1,2,3,4$).  Spherical scales $q=0$ and $q=1$ display the largest   aggregation  index CI$_{h}= exp(D_{q,h}^{R}),$  under the three  dependence models ($\theta =1, 100, 1/100$)  for all computed statistical distances. For this particular scenario where Log-Gaussian intensities are considered, the  log--intensity dependence range (reflected in parameter $\theta $), and the  statistical distance  chosen  (reflected in parameter $h$) only affect  the magnitude of the   distances computed at the first spherical scales. Specifically,  the clustering level, measured by the clustering  index  CI$_{h},$ increases when the dependence range becomes larger at these first scales  ($q=0,1,2$) around  the integer  values $h=1$ and $h=2$ of parameter $h.$

\begin{figure}[!htb]
\begin{center}
\includegraphics[height=4cm, width=4cm]{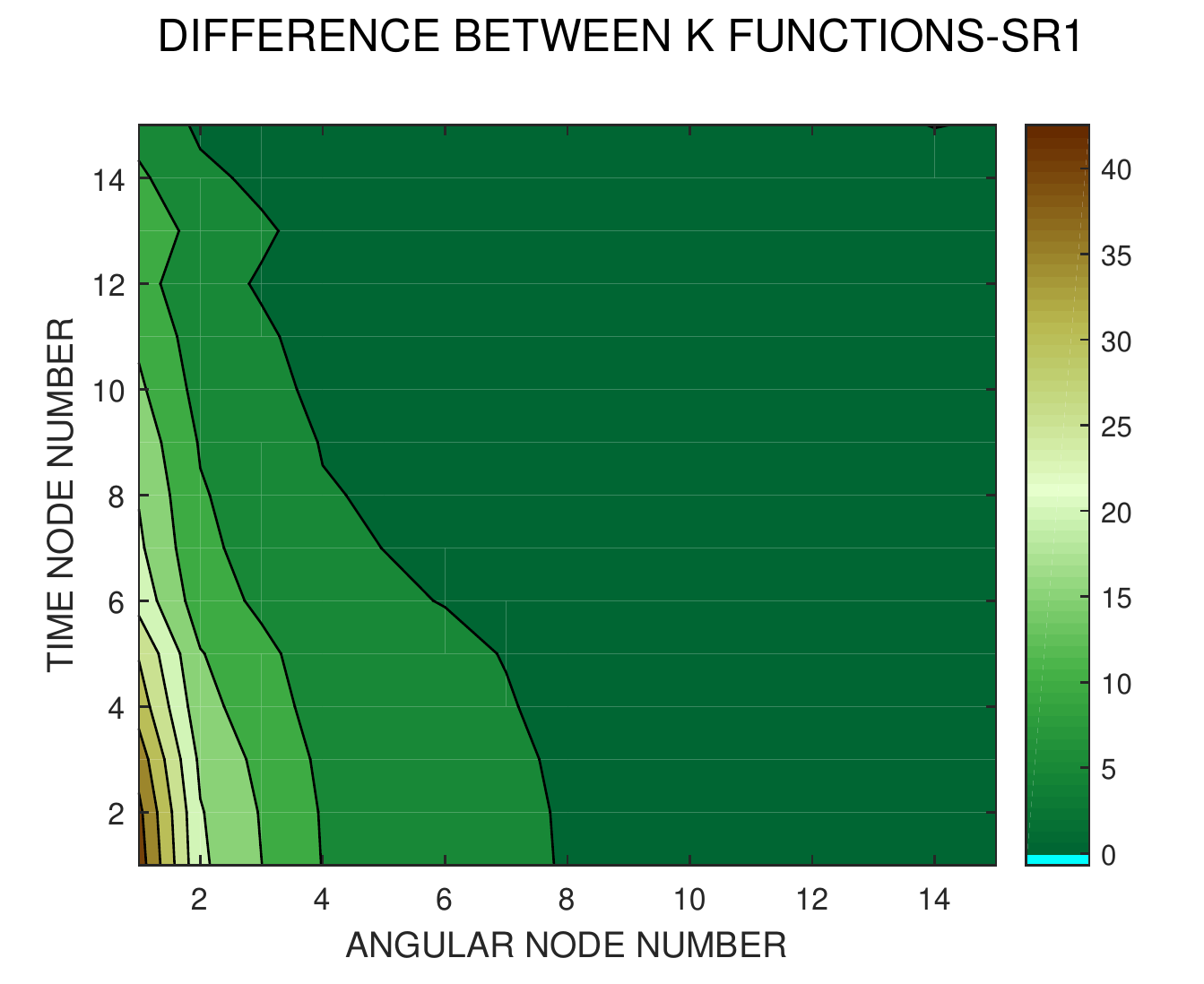}
\includegraphics[height=4cm, width=4cm]{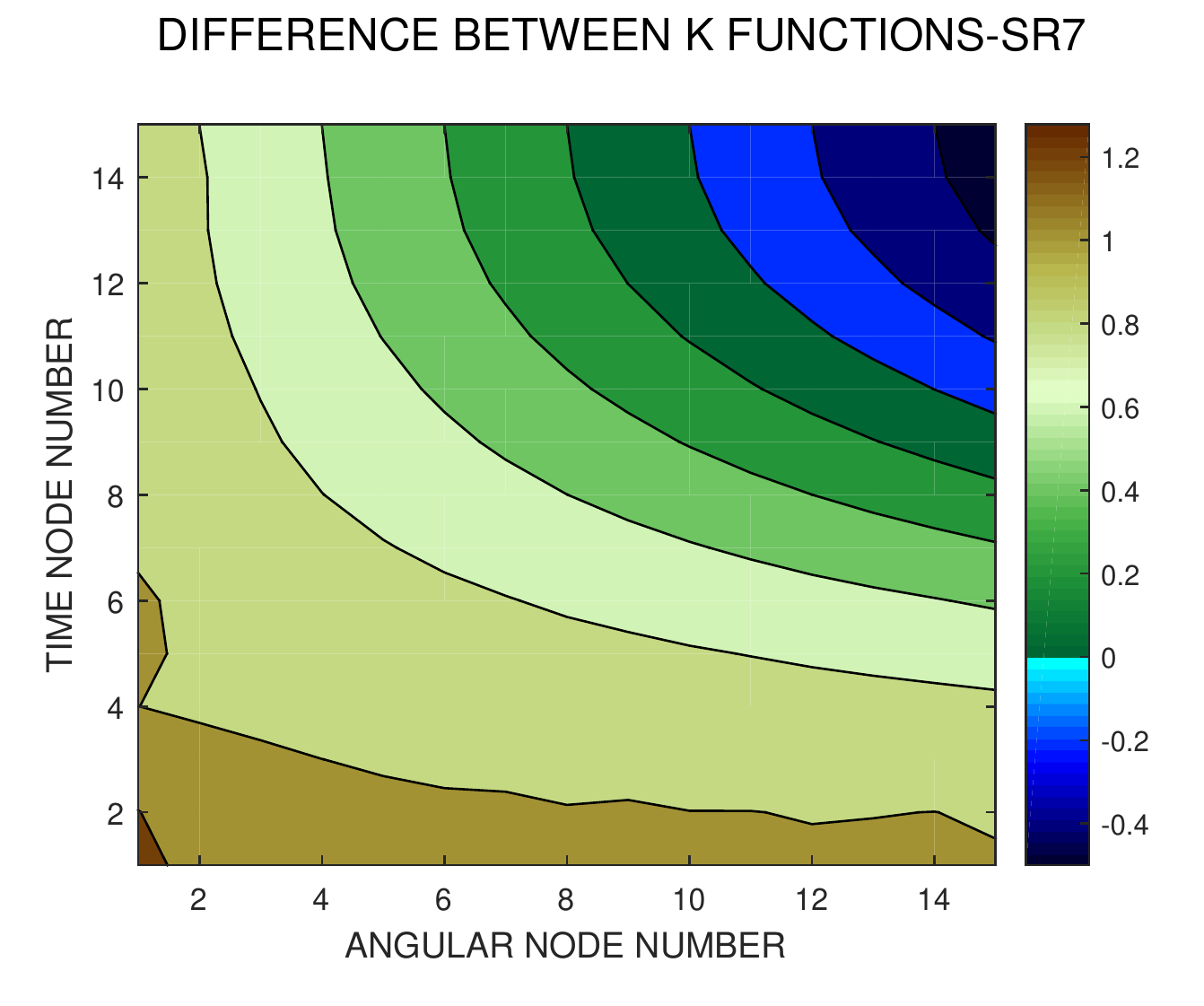}
\includegraphics[height=4cm, width=4cm]{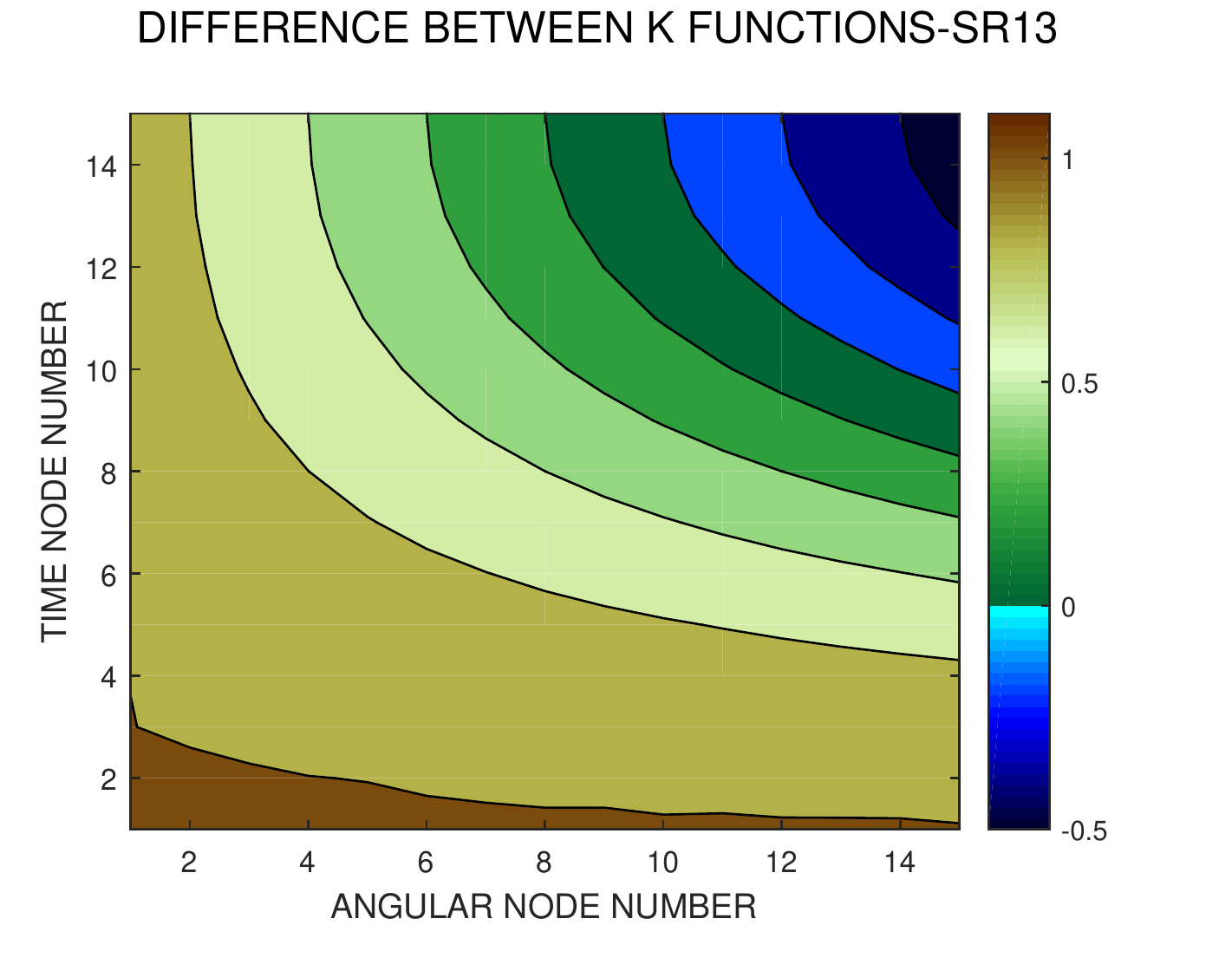}

\vspace*{0.5cm}

\includegraphics[height=4cm, width=4cm]{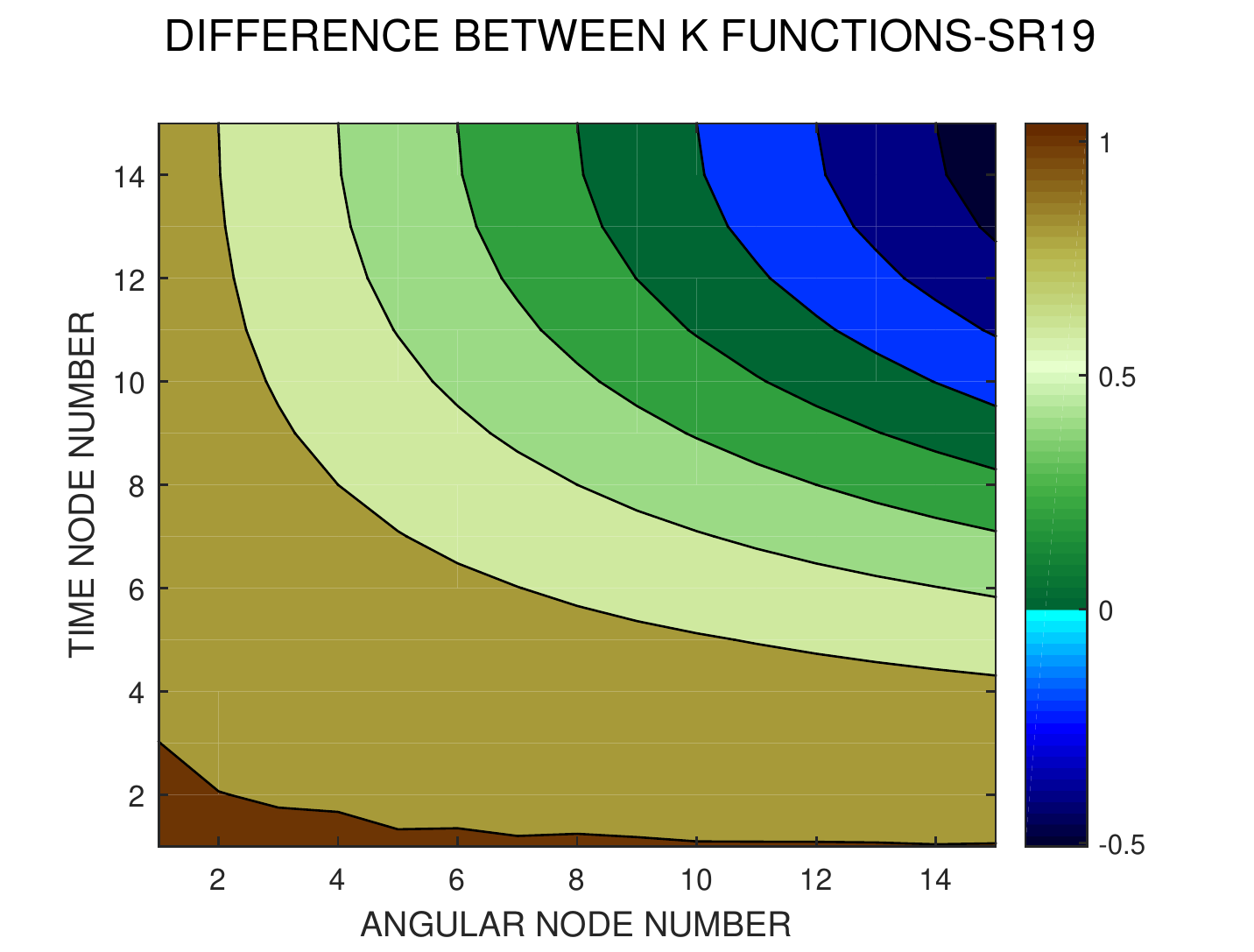}
\includegraphics[height=4cm, width=4cm]{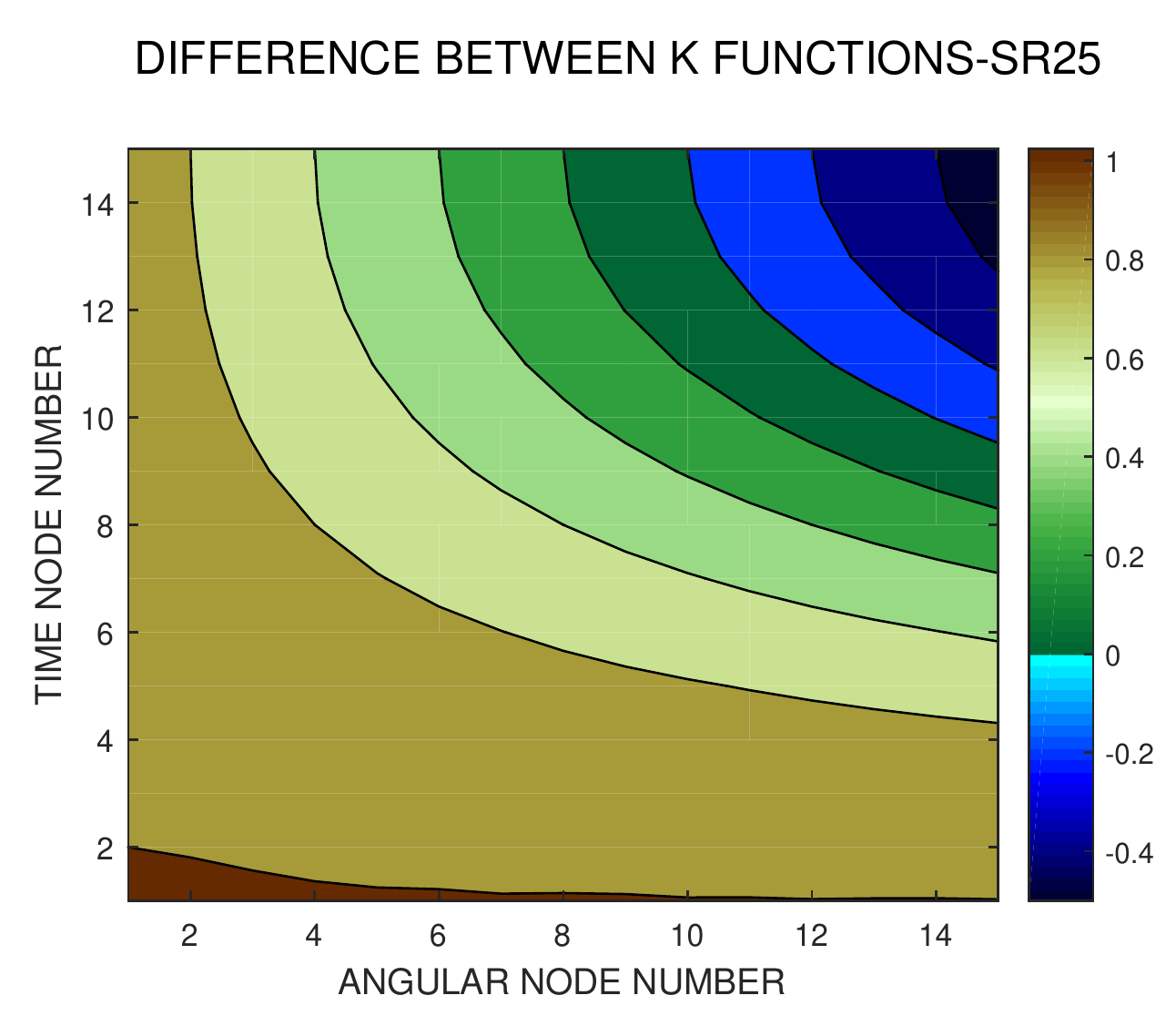}
\caption{ \scriptsize{\emph{Long--range dependence (LRD) Gaussian log--intensity}. Contour plots of pointwise values of empirical difference  $\widehat{K}_{q}-K_{\mbox{Pois}},$   for $q=1,7,13$ (top) and for $19, 25$ (bottom).  The generated spherical Log-Gaussian Cox process over the time interval $[0,10]$ has Legendre Fourier coefficients having  covariance function (\ref{temcovcoef}) with $\theta =1/100$ (LRD).
}}\label{fig3b}
\end{center}
\end{figure}

\begin{figure}[!htb]
\begin{center}
\includegraphics[height=4cm, width=4cm]{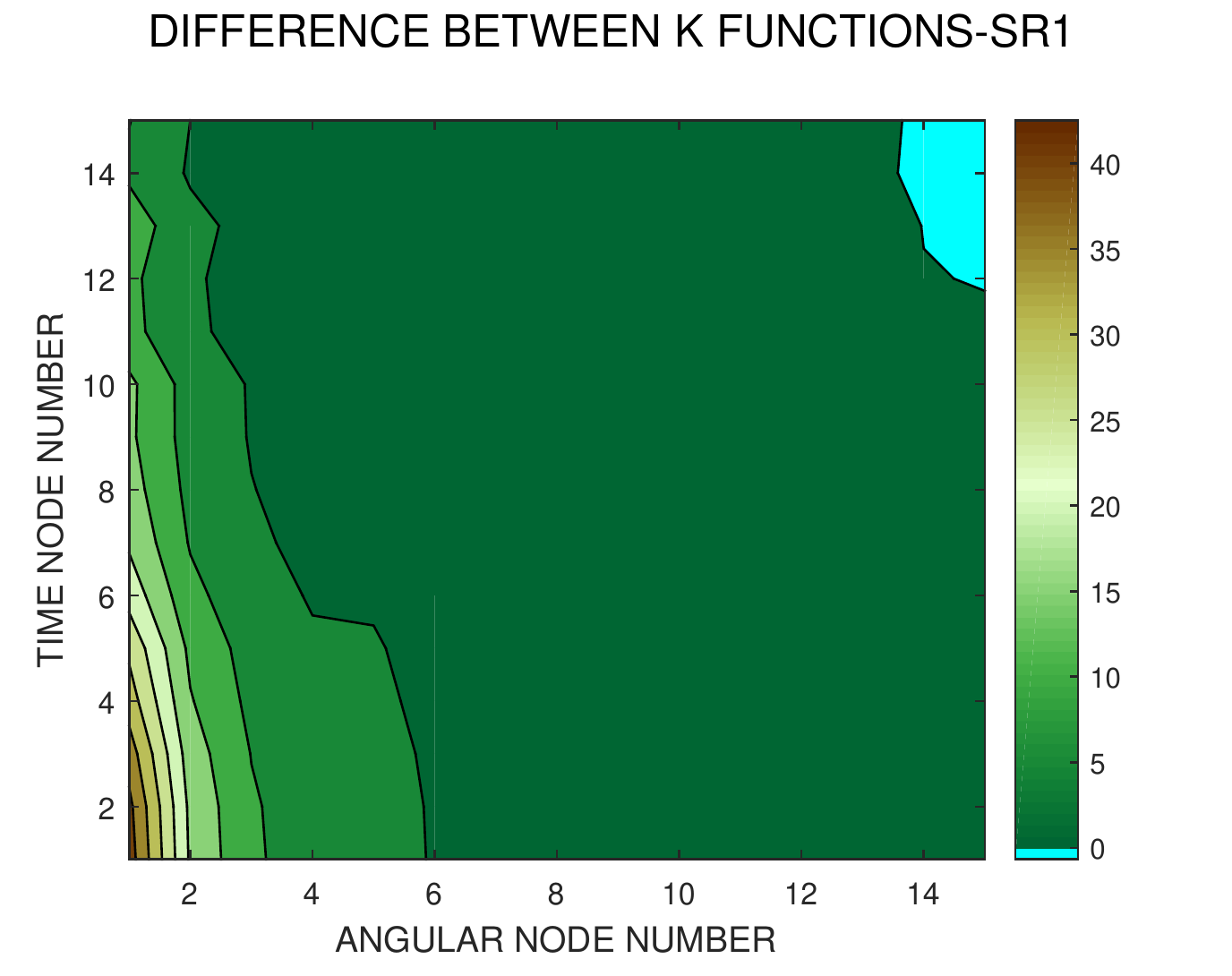}
\includegraphics[height=4cm, width=4cm]{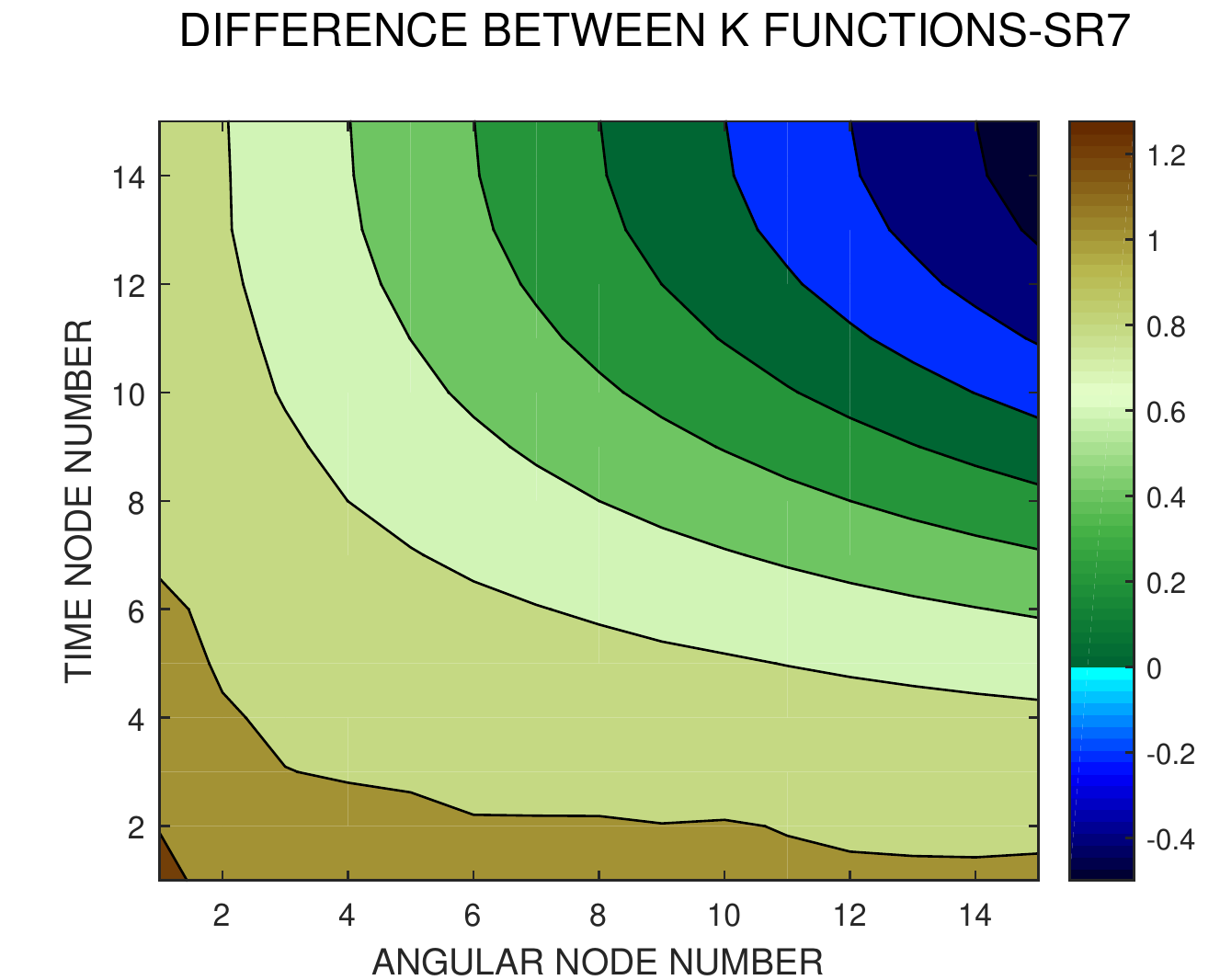}
\includegraphics[height=4cm, width=4cm]{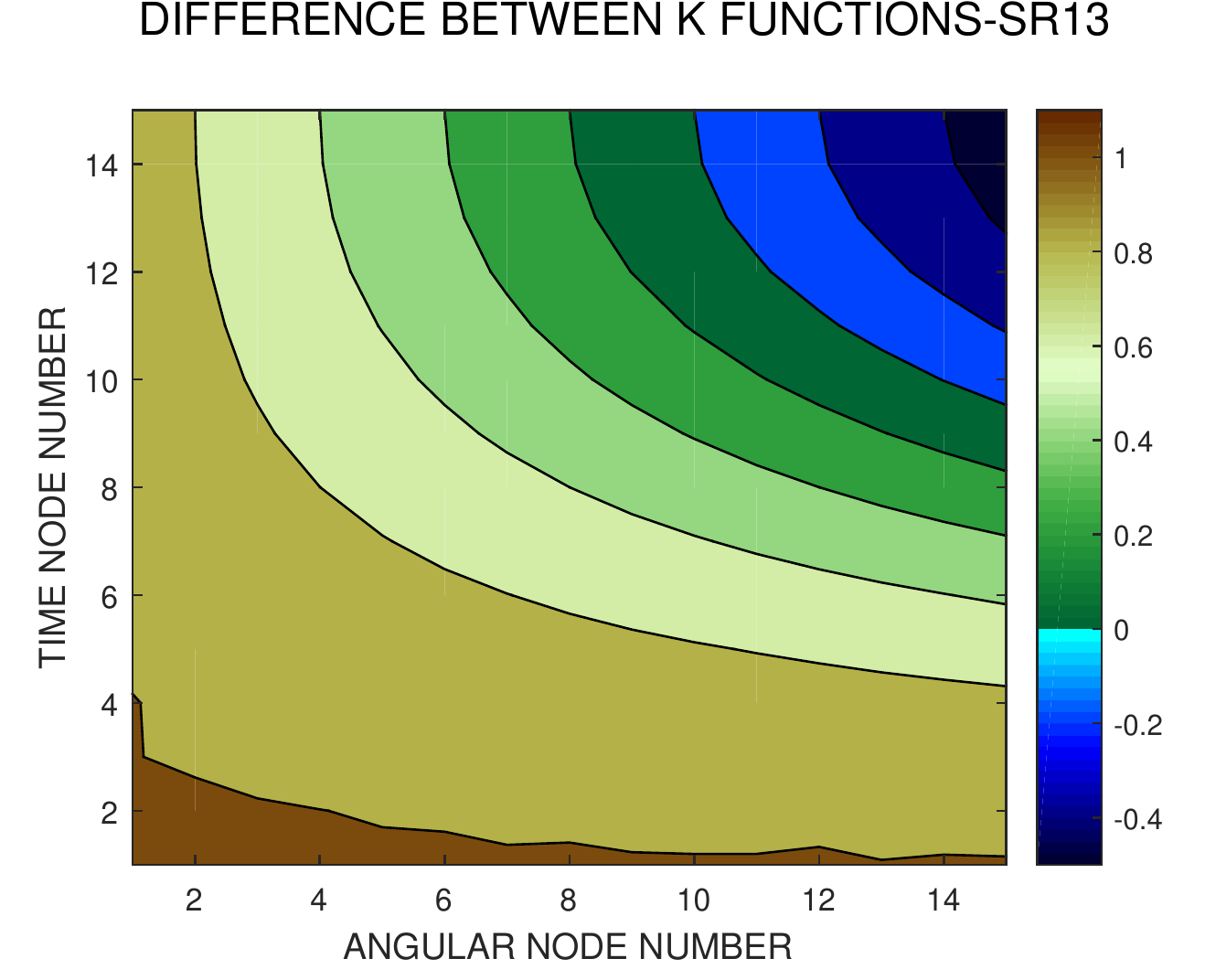}

\vspace*{0.5cm}

\includegraphics[height=4cm, width=4cm]{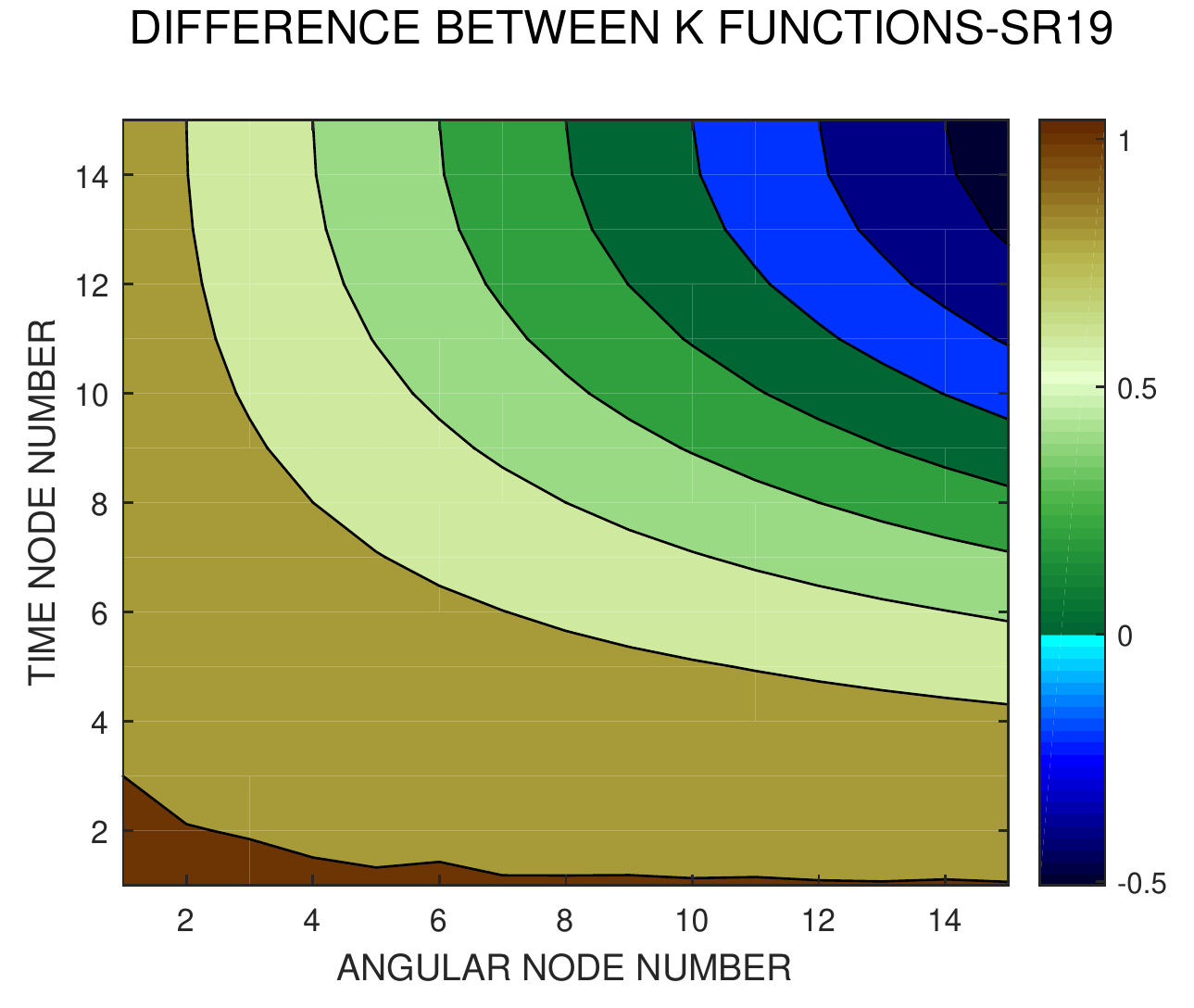}
\includegraphics[height=4cm, width=4cm]{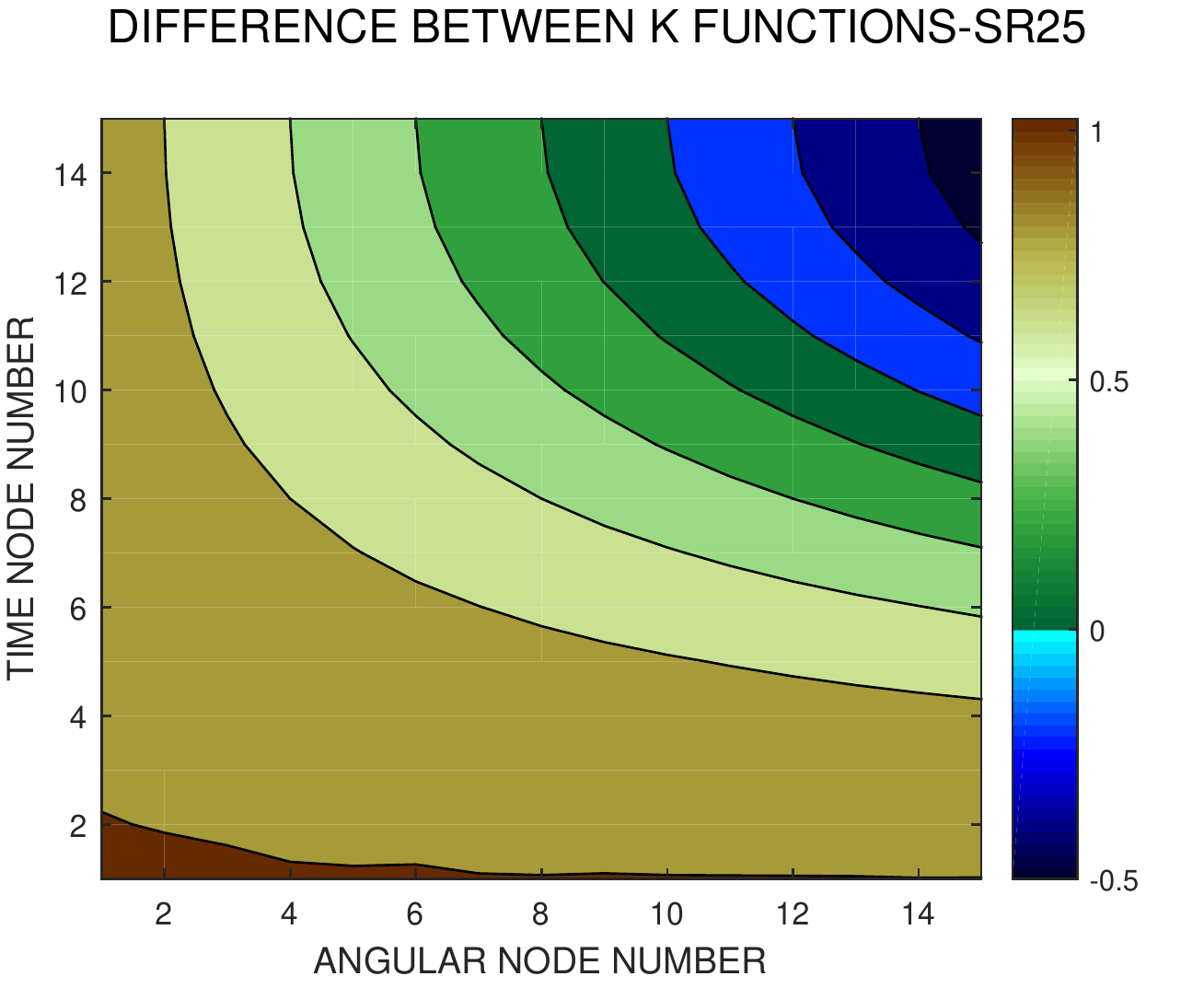}
\caption{ \scriptsize{\emph{Intermediate dependence range in the Gaussian log--intensity}. Contour plots of pointwise values of empirical difference  $\widehat{K}_{q}-K_{\mbox{Pois}},$   for $q=1,7,13$ (top) and for $19, 25$ (bottom). The  generated spherical Log-Gaussian Cox process over the time interval $[0,10]$ has Legendre Fourier coefficients having  covariance function (\ref{temcovcoef}) with $\theta =1.$
}}\label{fig3}
\end{center}
\end{figure}

\begin{figure}[!htb]
\begin{center}
\includegraphics[height=4cm, width=4cm]{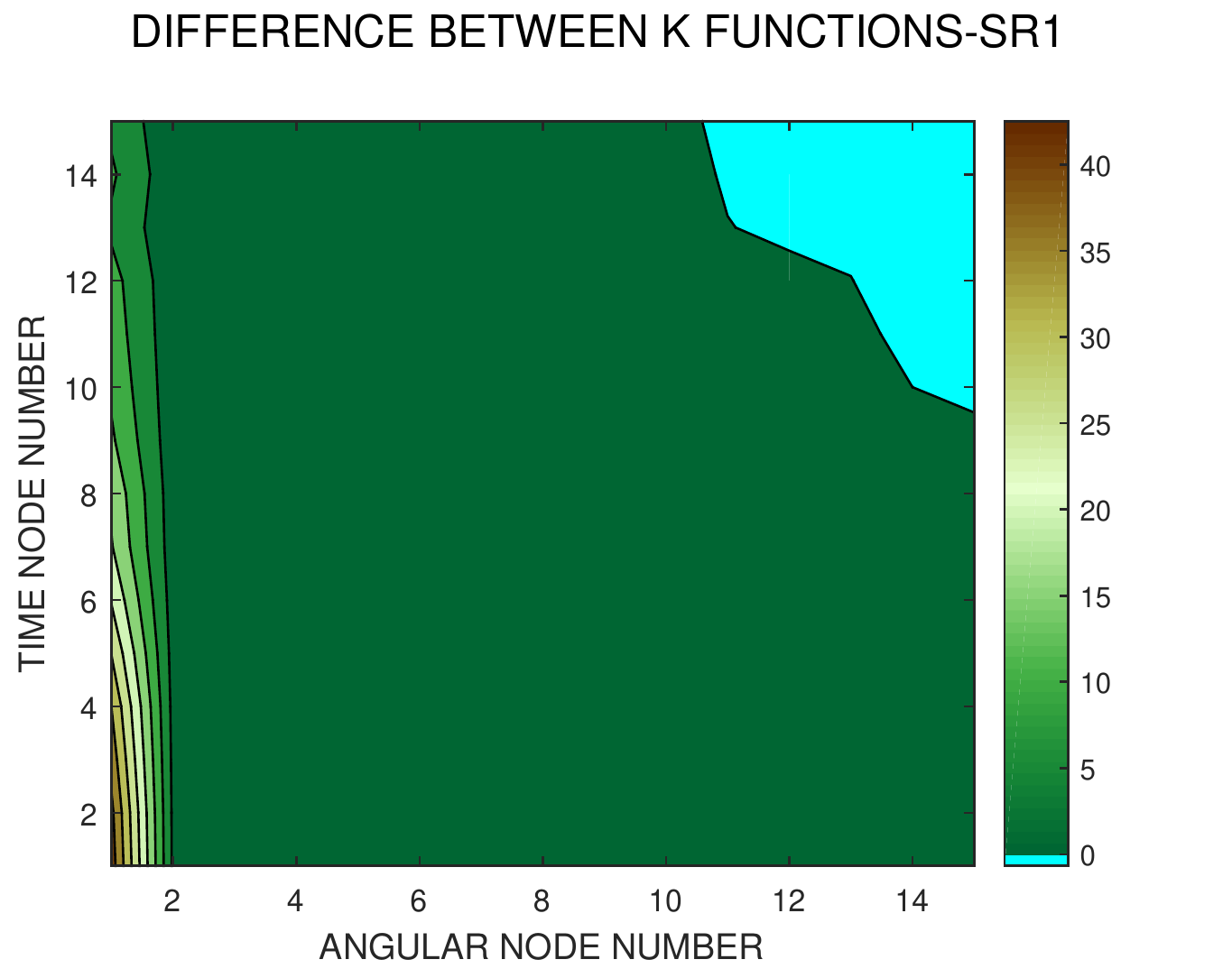}
\includegraphics[height=4cm, width=4cm]{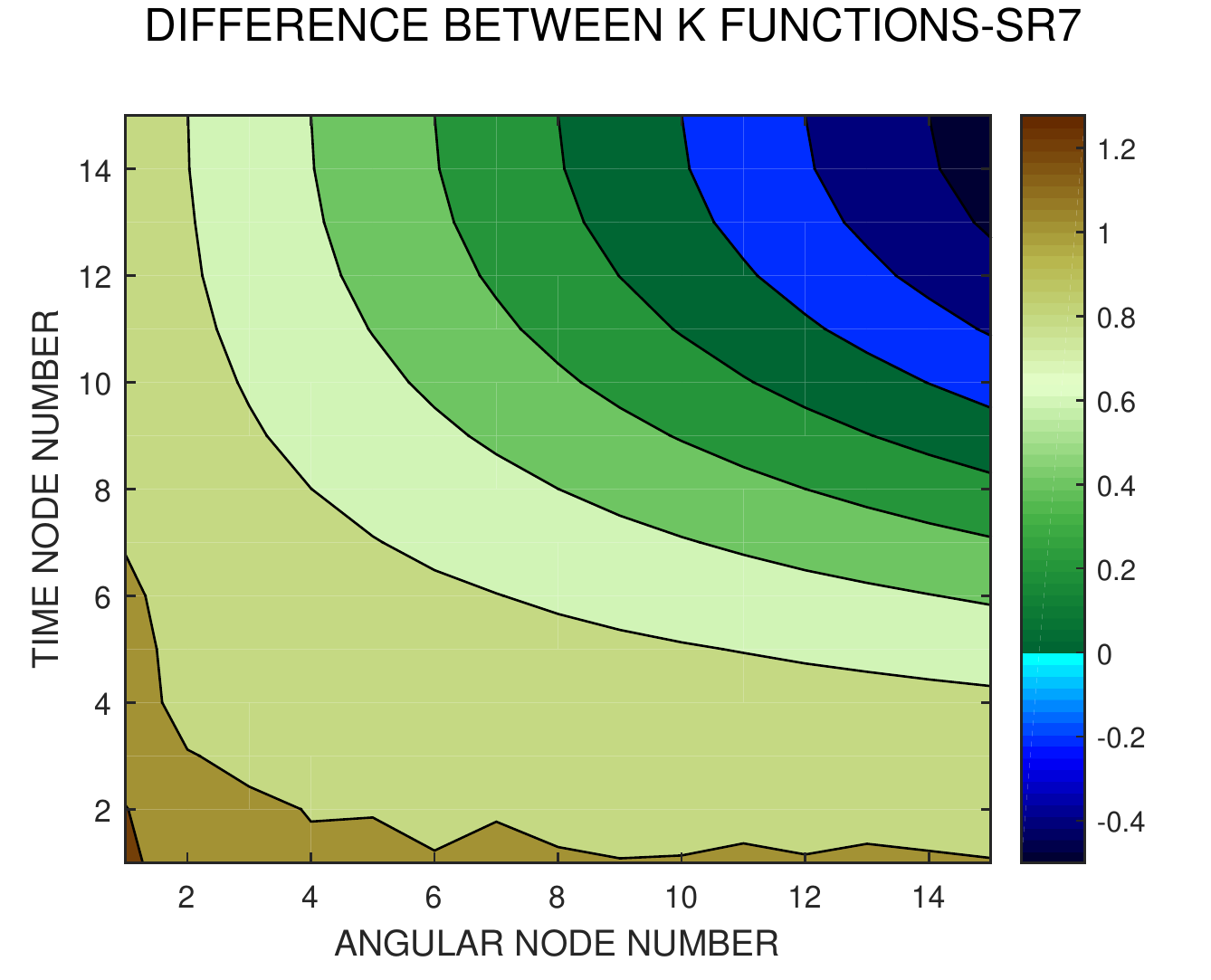}
\includegraphics[height=4cm, width=4cm]{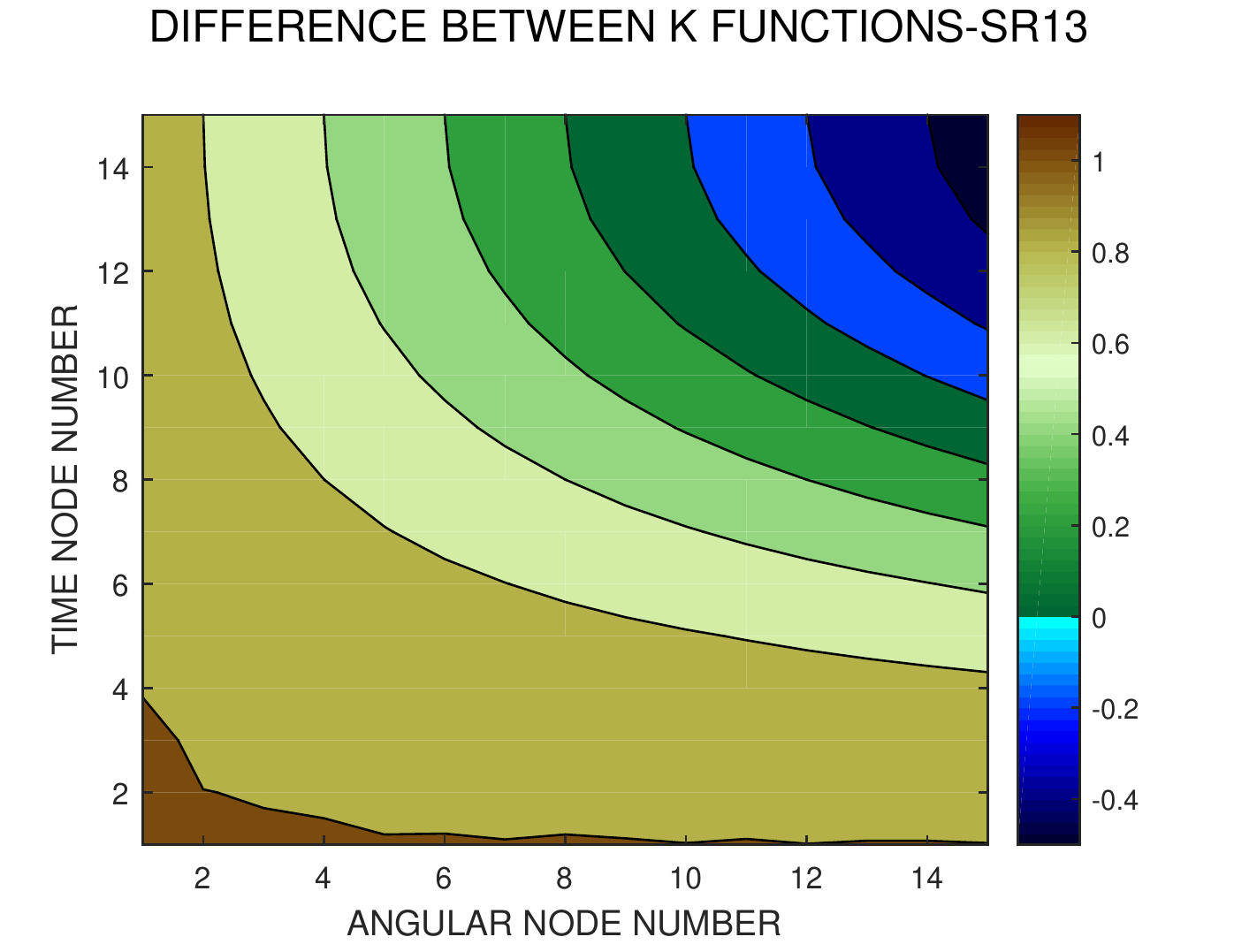}

\vspace*{0.5cm}

\includegraphics[height=4cm, width=4cm]{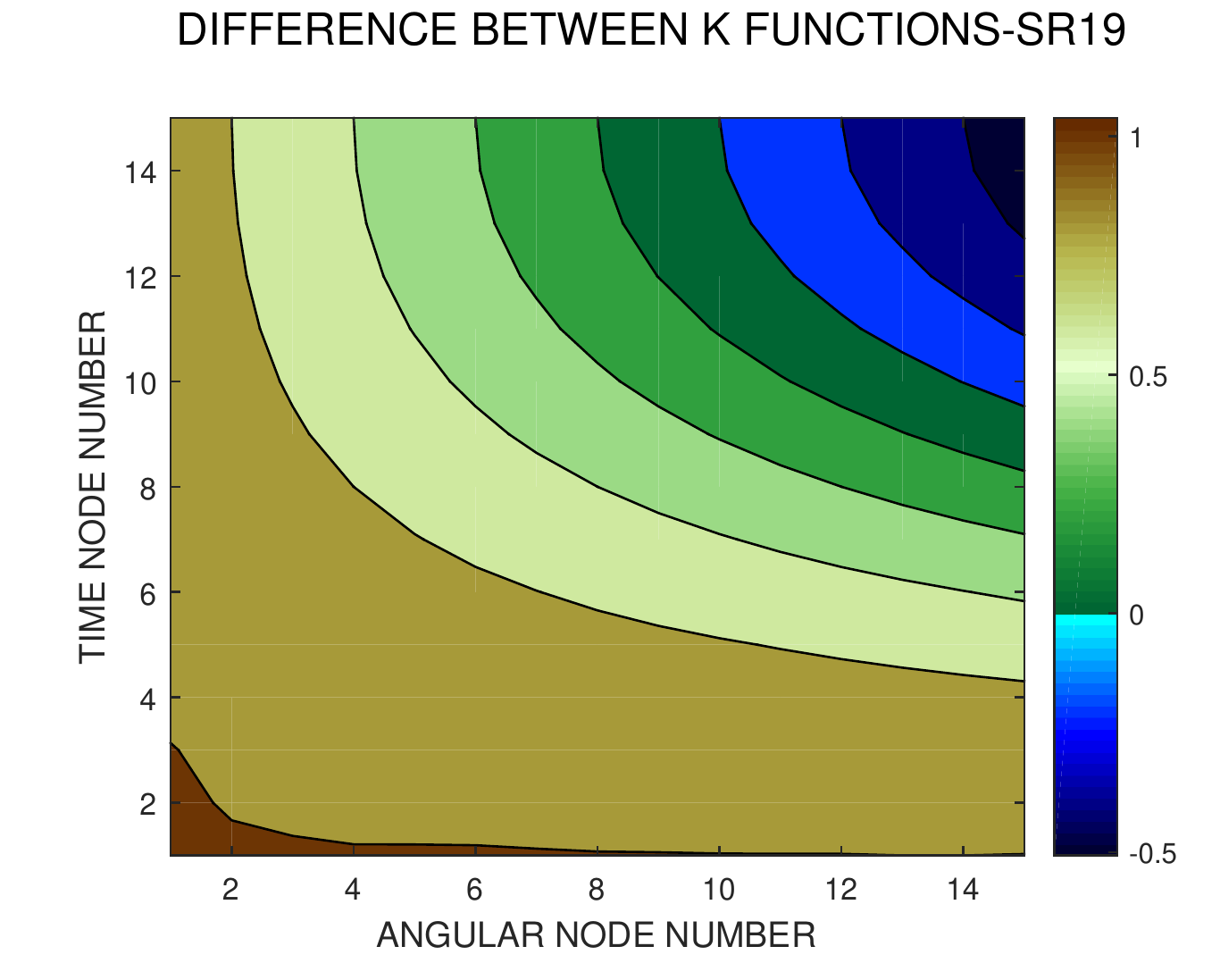}
\includegraphics[height=4cm, width=4cm]{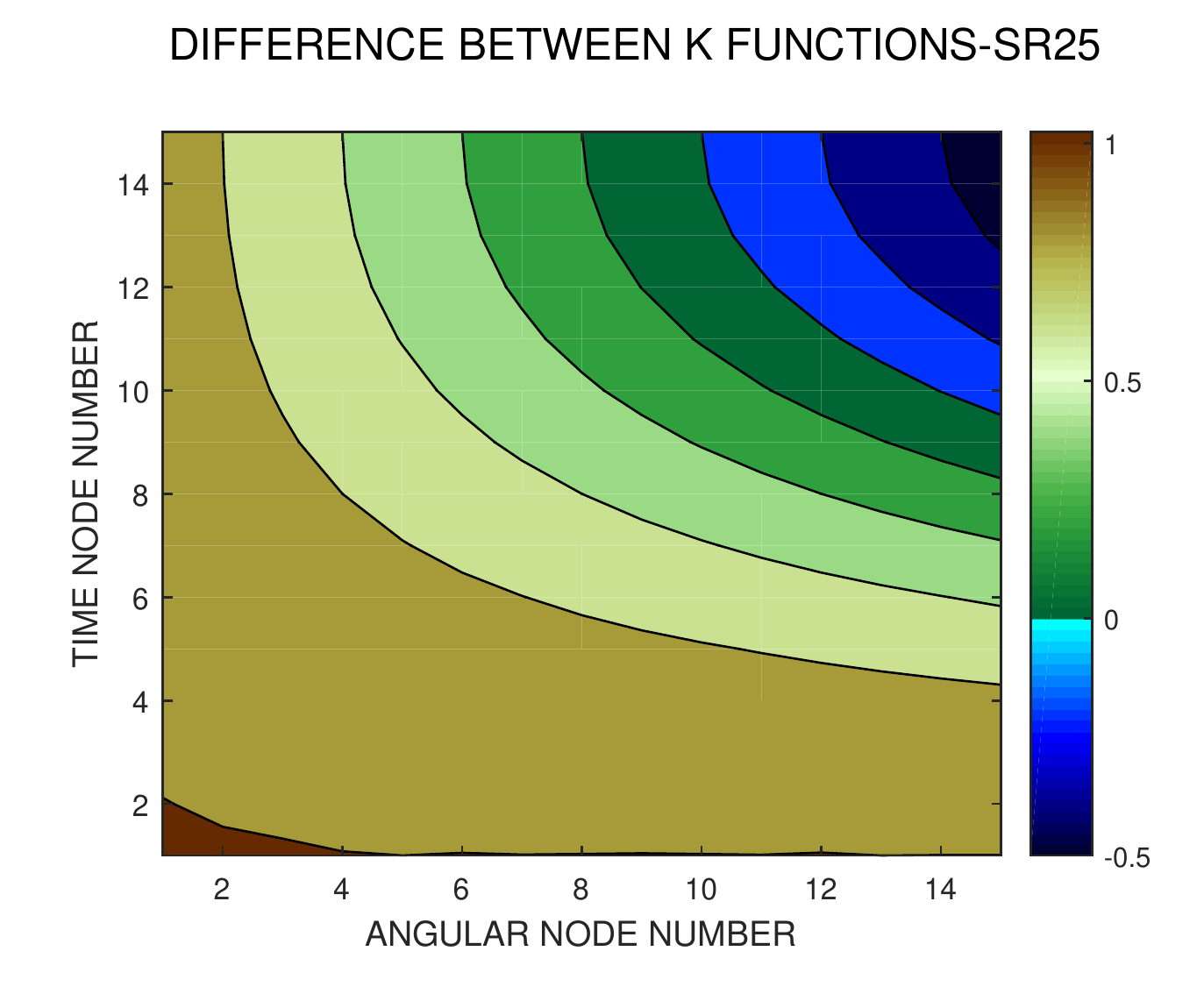}
\caption{ \scriptsize{\emph{Short--range dependence in the Gaussian log--intensity}. Contour plots of pointwise values of empirical difference  $\widehat{K}_{q}-K_{\mbox{Pois}},$   for $q=1,7,13$ (top) and for $19, 25$ (bottom). The  generated spherical Log-Gaussian Cox process over the time interval $[0,10]$ has Legendre Fourier coefficients having  covariance function (\ref{temcovcoef}) with $\theta =100.$
}}\label{fig3c}
\end{center}
\end{figure}

 Large  and small scale point pattern classification is here performed from Monte Carlo estimates $\widehat{K}_{q},$ $q=0,\dots,30,$ of functions $K_{q},$ $q=0,\dots,30,$  respectively.  The pointwise differences $\widehat{K}_{q}-K_{\mbox{\small Pois}},$ $q=0,\dots,30,$ with  $K_{\mbox{\small Pois}}$ denoting  as before the theoretical $K$ function of spatiotemporal spherical Poisson process, are plotted in Figures \ref{fig3b}--\ref{fig3c}, respectively corresponding to the long--, intermediate-- and short--range dependence cases of the log--intensity, for  Legendre scales  $q=1,7,13,19, 25.$  These functions are evaluated at the angular  distances $\{\theta_{i},i=1,\dots,14\}=\left\{0, 0.2244, 0.4488, 0.6732, 0.8976,\right.$ $\left. 1.1220, 1.3464, 1.5708,    1.7952,    2.0196,    2.2440,2.4684,    2.6928,    2.9172,    3.1416\right\},$\linebreak  in the interval $[0,\pi],$ and at the temporal distances $\left\{t_{i},i=1,\dots,14\right\}=$ \linebreak $\left\{0,   0.7143,    1.4286,    2.1429,    2.8571,    3.5714, 4.2857,    5.0000,    5.7143,    6.4286,7.1429,\right.$  \linebreak  $\left. 7.8571,    8.5714,    9.2857,   10.0000\right\},$  in the interval $[0,10].$

Figures \ref{fig3b}--\ref{fig3c} show that, for all log--intensity dependence ranges,   $\widehat{K}_{q}-K_{\mbox{{\small Pois}}}$ values
are decreasing when the Legendre spatial scale $q$ increases, going to zero when $q$ goes to infinity. Hence, for large values of $q,$ it can be observed that $\widehat{K}_{q}$ is pointwise approximating function $K_{\mbox{{\small Pois}}}$ (pointwise differences less than one for any temporal $t$ and angular $\theta $ distance values), supporting again  the computational results showed  in Figures  \ref{fig2}--\ref{fig2rde}. Thus,  regularity of spatiotemporal point patterns at   high Legendre  frequencies   ($q$ large) is observed, while aggregation or clustering is displayed at  low Legendre frequencies  ($q$ small), where bigger differences are induced by the log--intensity dependence range, i.e., larger positive pointwise discrepancies (stronger departure from regularity) are observed when the dependence range increases   (see, e.g., contour plots at the top-left  in Figures \ref{fig3b}, \ref{fig3} and \ref{fig3c}).

Summarizing, for small arguments $t$ and $\theta $ of $\widehat{K}_{q}-K_{\mbox{{\small Pois}}}$  functions,   more pronounced differences   are observed through Legendre scales,  while a  regular behavior is observed  for large values of $t$ and $\theta ,$ i.e., null values of  $\widehat{K}_{q}-K_{\mbox{{\small Pois}}}$ functions for every $q>1.$ For $q=1,$  positive pointwise discrepancies between $K$--functions hold for all ranges analyzed of $t$ and $\theta.$    One can also  observe that for this $q=1$ value the effect of the dependence range of the Gaussian log--intensity is stronger, increasing positive discrepancies between $K$ functions compared, for all arguments $t$ and $\theta.$  For the rest of scales ($q\geq 2$),  the effect  of the dependence range is more pronounced at small values of $t$ and $\theta .$
\section{Final comments}
\label{sec5}

Under stationarity in time and isotropy in space, the present paper performs a statistical analysis of point patterns on a connected and compact two--point homogeneous space. Specifically, this analysis is based on  Cox processes whose log--intensity (log--risk process)  is Gaussian or belongs to the class of second--order mean--square continuous elliptically contoured random fields  on a manifold (see, e.g., \cite{MaMalyarenko}).
In the Gaussian case, a countable  family of  independent stationary centered Gaussian processes defines the time--varying discrete Jacobi polynomial transform of the log--intensity  spatiotemporal  random field.  The $n$--order product  density then admits an expression in terms of the infinite--product of $n$--order product  densities  corresponding to  different Jacobi polynomial scales.

The  simulation study undertaken  is based on Monte Carlo numerical integration and least--squares parametric polynomial curve fitting, allowing the implementation of the proposed point pattern analysis, based on empirical statistical distances and $K$--functions, in terms of the time--varying discrete  Legendre polynomial transform. By exploiting the isometry properties with the  sphere,  the numerical results derived in this simulation study are extended to the  case of Log--Gaussian Cox processes on a connected and compact two point homogeneous space evolving time.
Thus,  one can conclude for the  wider introduced family of Log--Gaussian Cox processes, the regular behavior  (complete randomness) of the  point process  at
   large scales in the manifold.  While aggregation or clustering is displayed at small scale in the manifold. This Jacobi low-- and high-- frequency analysis is achieved at temporal and manifold micro--scale level  of the point pattern, by measuring the  statistical distance between  the $n$--order product  densities  of the analyzed point process, at different  Jacobi  polynomial scales, and the   $n$--order product density  of homogeneous Poisson process on the manifold over a time interval.   Different statistical distances are tested within the  Shannon-- and R\'enyi--entropy based distances.  The last ones providing a micro--scale aggregation (or clustering) index of the point pattern depending on  Jacobi scale. The effect of the temporal dependence range of the log--risk process is more pronounced at  low frequencies of discrete Jacobi polynomial transform. Particularly,  the  R\'enyi--based  micro--scale aggregation index increases when the temporal dependence range of the log--intensity increases at low Jacobi frequencies. While the effect of the temporal dependence range asymptotically  disappears at  high frequencies of the discrete Jacobi polynomial transform.
The   analysis of  point patterns in terms of the associated countable family of
    $K$--functions, arising from discrete Jacobi polynomial transform, also supports the conclusions of the micro--scale analysis based on statistical distances between  $n$--order product  densities.  Particularly,  at  low discrete  frequencies  (large  scale), stronger differences between  complete randomness and  scale--dependent $K$--functions of the analyzed point pattern  are observed.

The statistical methodology proposed for  analysis and multi--scale classification of point patterns  on a manifold over time,  in the context of connected and compact two--point homogeneous spaces,  within  the framework  of Cox processes, will be extended to the case of multifractal spherical log--risk processes  in a subsequent paper (see, e.g., \cite{Leonenko21}) .
Finally, we remark that the presented approach is applicable to further families of  point processes, including the family of determinantal point process  on a manifold  evolving time (see, e.g.,  \cite{Nielsen15}  and \cite{MollerRubak16} for the spatial spherical case).

\bigskip

\noindent \textbf{\small  Acknowledgements}
\noindent \scriptsize{This work has been supported in part by projects
MCIN/ AEI/PGC2018-099549-B-I00,  and CEX2020-001105-M MCIN/ AEI/10.13039/501100011033).}

\section*{References}

\bibliographystyle{plain}

\end{document}